\documentclass[aps,pra,showpacs,amssymb,nofootinbib,superscriptaddress,twocolumn,longbibliography]{revtex4-1}
\usepackage{epsfig,amsmath,amssymb,bm,epsf,graphicx,psfrag,bbm}
\usepackage{color, comment, verbatim}
\usepackage{ragged2e}
\usepackage{subfigure}
\usepackage{enumerate}
\usepackage{chngcntr}
\usepackage[T1]{fontenc}
\usepackage[english]{babel}
\usepackage[utf8]{inputenc}
\usepackage[colorlinks,breaklinks]{hyperref}
\usepackage{dsfont}
\usepackage{float}
\usepackage{tabularx}
\usepackage{graphicx}
\usepackage{amsbsy}
\usepackage{amsthm}

\usepackage{dsfont} 
\usepackage{tikz}
\usepackage[toc,page]{appendix}
\usepackage{titlesec}
\usepackage{xcolor}

\makeatletter
\newcommand\org@hypertarget{}
\let\org@hypertarget\hypertarget
\renewcommand\hypertarget[2]{%
  \Hy@raisedlink{\org@hypertarget{#1}{}}#2%
  }
\makeatother
\hypersetup{
	bookmarksnumbered,
	pdfstartview={FitH},
	citecolor={darkgreen},
	linkcolor={darkred},
	urlcolor={darkblue},
	pdfpagemode={UseOutlines}}
\definecolor{darkgreen}{RGB}{50,190,50}
\definecolor{darkblue}{RGB}{0,0,190}
\definecolor{darkred}{RGB}{238,0,0}

\usepackage[framemethod=tikz]{mdframed}
\definecolor{mycolor}{RGB}{78,121,200}
\definecolor{mycolor2}{RGB}{112, 48, 160}
\newmdenv[innerlinewidth=0.5pt, roundcorner=4pt,linecolor=mycolor,innerleftmargin=6pt,
innerrightmargin=6pt,innertopmargin=6pt,innerbottommargin=6pt]{mybox}

\usepackage{paracol}
\usepackage{tcolorbox}
\tcbset{colback=mycolor!5,colframe=mycolor,
fonttitle=\bfseries, float=htb}
\newtcolorbox[blend into=figures]{boxfigure}[3][]
{ float*=ht,width=\textwidth,lower separated=false, center upper,
title={#2},label= fig:#3,#1}
\newtcolorbox[blend into=figures]{smallboxfigure}[3][]
{float=ht,lower separated=false, blend before title=colon hang,
title={#2}, label= fig:#3 ,#1}
\newtcolorbox{smallbox}[3][]
{float=ht,lower separated=false, blend before title=colon hang,
title={#2}, label= fig:#3 ,#1}
\newtcolorbox[blend into=tables]{smallboxtable}[3][]
{float=t,lower separated=false, blend before title=colon hang,
title={#2}, label= table:#3 ,#1}
\newtcolorbox[blend into=tables]{bigboxtable}[3][]
{float*=t,lower separated=false, blend before title=colon hang, width = 2\linewidth,
title={#2}, label= table:#3 ,#1}
\newcolumntype{Z}{|>{\centering\arraybackslash}X}
\usepackage{enumerate}
\hypersetup{
	bookmarksnumbered,
	pdfstartview={FitH},
	citecolor={darkgreen},
	linkcolor={darkred},
	urlcolor={darkblue},
	pdfpagemode={UseOutlines}}
\definecolor{darkgreen}{RGB}{50,190,50}
\definecolor{darkblue}{RGB}{0,0,190}
\definecolor{darkred}{RGB}{238,0,0}
\usepackage{soul}

\newcommand{\be}{\begin{equation}}
\newcommand{\ee}{\end{equation}}
\newcommand{\ben}{\begin{equation*}}
\newcommand{\een}{\end{equation*}}
\newcommand{\bea}{\begin{eqnarray}}
\newcommand{\eea}{\end{eqnarray}}

\newcommand{\half}{\mbox{$\textstyle \frac{1}{2}$}}

\newcommand{\ket}[1]{\ensuremath{\left|\right.\!{#1}\!\left.\right\rangle}}

\newcommand{\bra}[1]{\ensuremath{\left\langle\right.\!{#1}\!\left.\right|}}
\newcommand{\braket}[2]{\ensuremath{\langle{#1}|{#2}\rangle}}
\newcommand{\ketbra}[2]{\ensuremath{|{#1}\rangle\langle{#2}|}}

\DeclareMathOperator{\sinc}{sinc}

\newcommand{\djj}{d\kern-0.4em\char"16\kern-0.1em}

\newcolumntype{s}{>{\hsize=.6\hsize}X}

\newcommand{\abs}[1]{\lvert #1 \rvert}

\begin{document}

\title{Characterising and Tailoring Spatial Correlations in Multi-Mode Parametric Downconversion}

\author{Vatshal Srivastav}
\email[Email address: ]{vs54@hw.ac.uk}
    \affiliation{Institute of Photonics and Quantum Sciences (IPAQS), Heriot-Watt University, Edinburgh, UK}

\author{Natalia Herrera Valencia}
    \affiliation{Institute of Photonics and Quantum Sciences (IPAQS), Heriot-Watt University, Edinburgh, UK}

\author{Saroch Leedumrongwatthanakun}
    \affiliation{Institute of Photonics and Quantum Sciences (IPAQS), Heriot-Watt University, Edinburgh, UK}
    
\author{Will McCutcheon}
    \affiliation{Institute of Photonics and Quantum Sciences (IPAQS), Heriot-Watt University, Edinburgh, UK}

\author{Mehul Malik}
    \email[Email address: ]{m.malik@hw.ac.uk}
    \affiliation{Institute of Photonics and Quantum Sciences (IPAQS), Heriot-Watt University, Edinburgh, UK}

\begin{abstract}
Photons entangled in their position-momentum degrees of freedom (DoFs) serve as an elegant manifestation of the Einstein-Podolsky-Rosen paradox, while also enhancing quantum technologies for communication, imaging, and computation. The multi-mode nature of photons generated in parametric downconversion has inspired a new generation of experiments on high-dimensional entanglement, ranging from complete quantum state teleportation to exotic multi-partite entanglement. However, precise characterisation of the underlying position-momentum state is notoriously difficult due to limitations in detector technology, resulting in a slow and inaccurate reconstruction riddled with noise. Furthermore, theoretical models for the generated two-photon state often forgo the importance of the measurement system, resulting in a discrepancy between theory and experiment. Here we formalise a description of the two-photon wavefunction in the spatial domain, referred to as the collected joint-transverse-momentum-amplitude (JTMA), which incorporates both the generation and measurement system involved. We go on to propose and demonstrate a practical and efficient method to accurately reconstruct the collected JTMA using a simple phase-step scan known as the $2D\pi$-measurement. Finally, we discuss how precise knowledge of the collected JTMA enables us to generate tailored high-dimensional entangled states that maximise discrete-variable entanglement measures such as entanglement-of-formation or entanglement dimensionality, and optimise critical experimental parameters such as photon heralding efficiency. By accurately and efficiently characterising photonic position-momentum entanglement, our results unlock its full potential for discrete-variable quantum information science and lay the groundwork for future quantum technologies based on multi-mode entanglement.
\end{abstract}
\maketitle

\section{Introduction}
\label{sec:Intro}

\begin{figure*}[ht!]
\centering\includegraphics[width=0.85\textwidth]{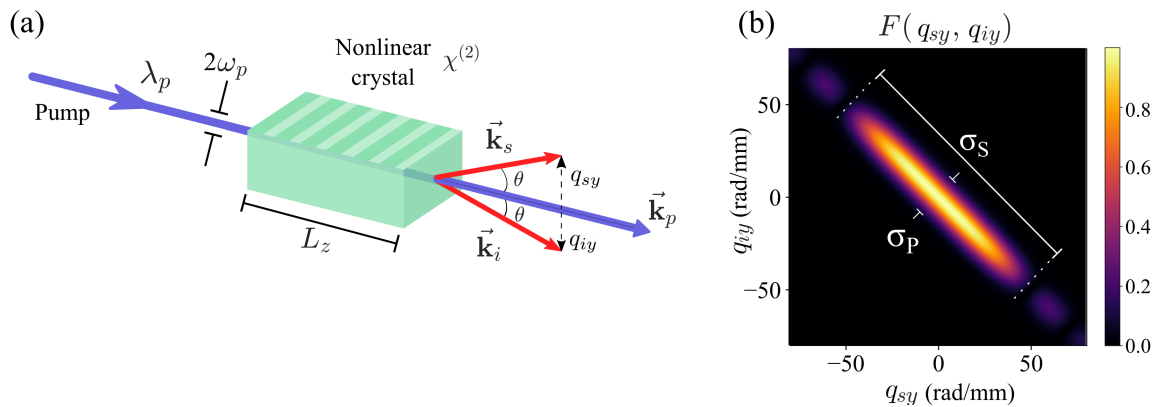}
\caption{(a) \textbf{Spontaneous parametric down conversion (SPDC)}: A nonlinear process where a coherent pump photon at wavelength $\lambda_P$ interacts with a nonlinear crystal ($\chi^{(2)}$) and generates two daughter photons (signal and idler) entangled in their position-momentum degrees-of-freedom. (b) \textbf{Joint-transverse-momentum-amplitude (JTMA)}: The density plot of a 2D slice of the JTMA $[F(q_{sx} = 0,q_{sy}, q_{ix} = 0,q_{iy})]$. The JTMA represents the transverse-momentum correlations between signal and idler photons and is characterized by two parameters, $\sigma_P$ and $\sigma_S$ (see Eq.~\eqref{eq:SPDCparameters}). The pump bandwidth parameter $\sigma_P$ is inversely propositional to the beam waist of the pump at the crystal plane ($w_P$) and dictates the strength of the correlations, while the generation bandwidth parameter $\sigma_S$ is determined by the length of the crystal $L_z$ and the wavelength of the pump $\lambda_P$, and determines the modal generation bandwidth of the JTMA.}
\label{fig:JTMA}
\end{figure*}

The Einstein, Podolsky, and Rosen (EPR) paradox lies at the heart of quantum mechanics~\cite{EPR1st:1935}. Using the paradigmatic example of two quantum particles sharing perfect correlations (or anti-correlations) between their complementary properties of position and momentum, EPR postulated an inconsistency between local realism and the completeness of quantum mechanics~\cite{Reid:2009us}. A physical realisation of the original EPR experiment proved challenging, and much of the subsequent theoretical and experimental work focused on a discrete version of the EPR paradox postulated by Bohm and formalised by Bell's inequality~\cite{Bohm1957,Bell1964,Brunner:2014}. While discrete-variable experiments such as ones based on polarisation~\cite{Clauser:1969ff,Aspect1982} have laid the foundation for the quantum technologies of today, the exploration of continuous quantum properties in the vein of the original EPR gedankenexperiment has recently flourished~\cite{Shin2019, Wasak2018, Keller2014, Ndagano2020, Edgar:2012, Moreau2012}, thanks to a series of experimental advances and several practical motivations.

Pairs of photons produced in nonlinear spontaneous parametric downconversion (SPDC) provide a natural platform for tests of EPR entanglement. Photons generated in SPDC are correlated/anti-correlated in their position and momentum owing to the conservation of energy and momentum that governs this process~\cite{BoydNLO,HongMandel85,Malygin1985,Schneeloch:2016ch,Brambilla2010,Walborn:2010tj}. While this source was adapted for the earliest violations of Bell's inequality based on discrete-variable polarisation entanglement, the ability to harness its inherent position-momentum correlations has led to a recent explosion of interest in high-dimensional entanglement of photonic spatial modes~\cite{Erhard:2019ux,Friis:2019hg,Bavaresco:2018gw}---ranging from demonstrations of high-dimensional Bell-like inequalities~\cite{Dada:2011dn}, composite quantum state teleportation~\cite{Wang:2015dm}, to exotic forms of multi-photon entanglement~\cite{Malik:2015we,Erhard:2018iua,Hiesmayr2016, Hu2020}. High-dimensional (qudit) entanglement also provides significant advantages over qubit-based systems in the form of increased information capacity~\cite{Valencia:2020gx,Cao:2020,Cozzolino:2019ct,Erhard:2017gl,Steinlechner:2017bw,Mafu:2013jk} and robustness to noise~\cite{Ecker-Huber2019,Zhu:2019tb,Marcin2013, Vertesi:2010bq}, making it a very promising platform for next-generation quantum technologies such as device-independent quantum cryptography~\cite{Acin:2007db}. Thus, the ability to efficiently and accurately characterise the underlying two-photon state entangled in its continuous position-momentum degrees-of-freedom is of paramount importance.

Modelling the generation of entangled photons in a continuum of modes allows for the identification of the effective number of entangled modes that are present in the system~\cite{Miatto:2012ck,Schneeloch:2016ch}---the so-called \textit{generation bandwidth}. In addition, precise knowledge of the continuous position-momentum correlations is crucial for accurately tailoring spatial mode bases that maximise metrics relevant to discrete-variable quantum information processing, such as the entanglement-of-formation ($E_{\textrm{oF}}$), entanglement dimensionality, and the state fidelity. Experimental reconstruction of a position-momentum entangled state presents some unique challenges---detector technology limits one to scanning through the position/momentum space of interest with a single-mode detector, which inherently introduces loss and involves very long measurements times~\cite{Howell:2004fc,Schneeloch:2019uw}. Recent work has pushed the capabilities of arrayed single-photon detectors to reconstruct such states faster~\cite{Edgar:2012, Moreau2012,Defienne:2018bi,Ndagano2020}, however these techniques still suffer from resolution limits, loss, and an associated large noise background.

In this work, we formulate a theoretical model for a two-photon position-momentum entangled state---the collected joint-transverse-momentum-amplitude (JTMA)---that incorporates the generation as well as the measurement system used in an experiment. We propose and demonstrate a practical and fast method to fully characterise the collected JTMA using a simple $\pi$-phase step scan, akin to a classical knife-edge measurement of a laser beam. Our method, known as the $2D\pi$-measurement, allows us to measure the state parameters independent of our knowledge of the optical system and crystal properties. While here we implement our measurement technique with programmable phase-only spatial light modulators (SLMs), its simplicity enables it to be performed with low-cost components such as a microscope glass slide. We demonstrate the versatility of our measurement scheme by implementing it on two experiments in the continuous-wave near-infrared and pulsed telecom wavelength regimes. Finally, we discuss how accurate knowledge of the collected JTMA enables us to generate tailored discrete-variable high-dimensional entangled states that maximise a desired property such as entanglement-of-formation or entanglement dimensionality, or optimise experimental measures such as photon heralding efficiency. Our methods have significant potential implications for entanglement-based quantum technologies as well as fundamental tests of quantum mechanics, and can be translated to other continuous degrees of freedom such as time-frequency in a straightforward manner.

\section{Theory}
\label{sec:Theory}
\subsection{Collected Bi-photon JTMA}
As shown in Fig.~\ref{fig:JTMA}a, the process of spontaneous parametric downconversion (SPDC) results in the generation of a two-photon state whose correlations in momentum space can be well approximated by a function that we call the \textit{joint-transverse-momentum-amplitude} (JTMA). In Appendix~\ref{sec:JTMAapp}, we derive the full JTMA of the two-photon wavefunction produced in Type-II SPDC by a periodically poled nonlinear crystal designed to achieve phase-matching at degenerate frequencies in the collinear configuration. At degenerate frequencies and imposing a Gaussian transverse pump profile across the crystal, we show how it approximates the well-known form~\cite{Schneeloch:2016ch,Miatto:2012ck}:

\be
F(\textbf{q}_s, \textbf{q}_i)= \mathcal{N}_1\underbrace{\exp\bigg( \frac{-|\textbf{q}_s+\textbf{q}_i|^2}{2\sigma_P^2}\bigg)}_{\textrm{Pump profile}}\times \underbrace{\text{ sinc}\bigg(\frac{|\textbf{q}_s - \textbf{q}_i|^2}{\sigma_S^2}\bigg)}_{\textrm{Phase-matching condition}},
\label{eq:JTMAformula}
\ee
where $\textbf{q}_s (\textbf{q}_i)$ is the transverse-momentum wave-vector for the signal (idler) photon and $\mathcal{N}_1$ is a normalization constant.
The first term in Eq.~\eqref{eq:JTMAformula} represents the transverse wave-vector components of the pump, while the second term represents the phase-matching condition imposed on the down-conversion process by the nonlinear crystal. 
The pump width parameter $\sigma_P$ depends on the $1/e^2$ beam radius of the pump's intensity profile at the crystal plane ($w_p$), while the generation width parameter $\sigma_S$ is determined by the crystal length $L_z$ and pump wavevector inside the crystal $\textit{k}_p$. These parameters are defined as~\cite{Miatto:2012bv}: 
\be
\label{eq:SPDCparameters}
\sigma_P =\frac{\sqrt{2} }{w_p}, \qquad \sigma_S = \sqrt{\frac{4\textit{k}_p}{L_z}}.
\ee
The pump wavevector $\textit{k}_p  = n_P\tfrac{2 \pi}{\lambda_P}$, where $n_P$ is the refractive index of the nonlinear crystal at pump wavelength $\lambda_P$.
The interplay between the parameters $\sigma_P$ and $\sigma_S$ determines the momentum correlations between the signal and idler photon.
In the case where $\sigma_P<\sigma_S$, the parameter $\sigma_P$ dictates the strength of the momentum correlations, whilst $\sigma_S$ represents the generation width of the JTMA function (see Fig.~\ref{fig:JTMA}b).
For instance, a very broad (plane wave) pump beam has $\sigma_P \ll \sigma_S$, and the approximation $\sigma_P\rightarrow 0$ can be made. Here, the pump contribution to the JTMA approximates to $\delta(\textbf{q}_s + \textbf{q}_i)$, which results in perfect anti-correlations between the signal and idler transverse momenta. In contrast, for a non-zero $\sigma_P$, the degree of correlations decreases as the $\sigma_P$ increases.
For very thin crystals ($L_z \rightarrow 0$), the generation width of momentum correlation tends to infinity ($\sigma_S \rightarrow \infty$).
However, as shown in~\cite{Baghdasa2021}, this approximation breaks for small beam waist $w_p$, revealing  the importance of a finite value of $\sigma_S$ and $\sigma_P$ in reality.
\begin{figure}[t!]
\centering\includegraphics[width=1\columnwidth]{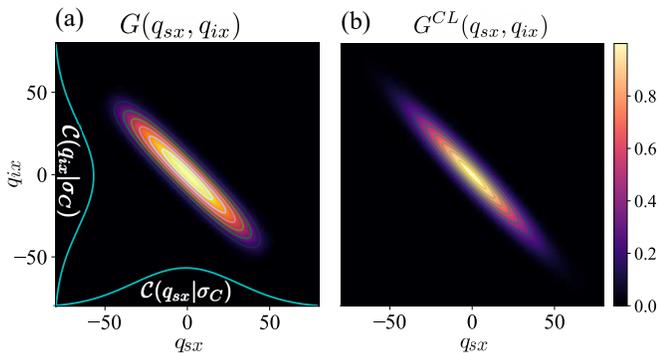}
\caption{\textbf{Collected JTMA}: (a) The density plot shows a 2D slice of the collected JTMA $[G(q_{sy} = 0,q_{sx}, q_{iy} = 0,q_{ix})]$ from Eq.~\eqref{eq:collectedJTMA}, where the Gaussian distributions on the axes $q_{sx}$ and $q_{ix}$ represent the collection modes $\mathcal{C}(\textbf{q}_n|\sigma_C)$ that suppress the Sinc sidelobes of the generated JTMA $F(\textbf{q}_s, \textbf{q}_i)$ shown in Fig.~\ref{fig:JTMA}b. (b) Collected JTMA under the \textit{collection-limited} (CL) approximation, which is valid if $\sigma_S \gtrapprox \sqrt{2}\sigma_C$, resulting in the double-Gaussian form shown in Eq.~\eqref{eq:CLapprox}.}
\label{fig:Coll_JTMA}
\end{figure}

Thus far, we have only discussed the generated two-photon state, which depends solely on the pump and crystal parameters $\sigma_P$ and $\sigma_S$. However, the measured momentum correlations also depend on the configuration of the detection system. Photonic spatial modes are routinely measured in the laboratory via a combination of a holographic spatial light modulator (SLM) and single-mode fibre (SMF) that together act as a spatial-mode filter~\cite{Qassim:2014fp,Bouchard:2018hr}. To model the effects of such a spatial-mode filter on the JTMA, we first consider the effect of phase-only SLMs placed in the Fourier plane of the nonlinear crystal. By implementing a diffractive hologram, an SLM can be used to apply an arbitrary amplitude and phase function $\Phi^{(x)}(x)$ on an incident light field, which is then projected onto the Gaussian single-mode of the collection fibre. Collection modes have been considered in the context of spatial-mode entanglement \cite{Miatto:2012ck}. For convenience, we choose to work in momentum space at the crystal plane so the collection mode takes the form $\mathcal{C}(\textbf{q}|\sigma_C)=(\sqrt{\pi}\sigma_C)^{-\half}\exp\{- \tfrac{|\textbf{q}|^2}{\sigma_C^2}\}$, characterised by a collection bandwidth $\sigma_C$ in transverse angular momentum units. However, since the collection mode commutes with the SLM holograms, here we incorporate it into the state itself. We can then consider the \textit{collected} bi-photon JTMA, which allows us to model all the properties of the state that we can access and manipulate with SLM holograms. Since the collection mode:
\bea
G(\textbf{q}_s, \textbf{q}_i)&=&\mathcal{C}(\textbf{q}_s|\sigma_C)\times\mathcal{C}(\textbf{q}_i|\sigma_C)
\times F(\textbf{q}_s, \textbf{q}_i)\nonumber \\
&=&\tfrac{\mathcal{N}_1}{\sqrt{\pi}\sigma_C}\underbrace{\exp\bigg(- \frac{|\textbf{q}_s|^2}{\sigma_C^2}\bigg)
\exp\bigg(- \frac{|\textbf{q}_i|^2}{\sigma_C^2}\bigg)}_{\text{Collection widths}}\label{eq:collectedJTMA}\\&\times&
\underbrace{\exp\bigg( \frac{-|\textbf{q}_s+\textbf{q}_i|^2}{2\sigma_P^2}\bigg)}_{\text{Correlations strength}} \times \underbrace{\text{sinc}\bigg(\frac{|\textbf{q}_s - \textbf{q}_i|^2}{\sigma_S^2}\bigg)}_{\text{Generation width}} .
\nonumber\eea
The choice of $\sigma_C$ (relative to $\sigma_S$ and $\sigma_P$) limits the entanglement dimension and collection efficiency that can be achieved. In practice, $\sigma_C$ can be carefully set through a choice of the optical system parameters (discussed in detail in Section.~\ref{sec:exp}). 
As can be seen in Fig.~\ref{fig:Coll_JTMA}a, the effect of including Gaussian collection modes with a specific $\sigma_C$ suppresses the Sinc sidelobes of the generated JTMA (Fig.~\ref{fig:JTMA}b). We can then write the collected two-photon coincidence (joint) probability of detecting the signal and idler photons when displaying the hologram functions $\Phi_s$  and $\Phi_i$ respectively, as
\be
\textbf{Pr}(\Phi_s,\Phi_i)=\biggl| \int d^2\textbf{q}_s \int d^2 \textbf{q}_i \Phi_s(\textbf{q}_s) \Phi_i(\textbf{q}_i) G(\textbf{q}_s, \textbf{q}_i)\biggr|^2. \label{eq:probcon}
\ee
\subsection{$2D\pi$-Measurement}
To characterize the collected JTMA of a two-photon state, one requires to know the parameters $\sigma_P$, $\sigma_S$, and $\sigma_C$. While these parameters can be calculated from the optical system properties, one would practically need to be able to measure them independently and verify that an experimentally generated state is indeed close to what theory predicts. Additionally, this capability is of particular relevance when the optical system is complex, unknown, or inaccessible.

Here, we introduce a simple measurement scheme that we call the $2D\pi$-measurement, which allows us to estimate the parameters $\sigma_P$ and $\sigma_C$, and obtain an accurate two-photon JTMA. This measurement is related to the classical knife-edge measurement routinely used to measure the transverse profile of a laser beam~\cite{Arnaud:71}. The $2D\pi$-measurement can be thought of as a two-photon phase-only knife-edge, where a $\pi$-phase step is scanned across both the signal and the idler photons, resulting in a 2D function containing information about the two-photon JTMA. The $\pi$-phase step is easily implemented via phase-only spatial light modulators (SLMs) placed in each path. The choice of using a phase edge over amplitude prevents high photon loss during the measurement, making the $2D\pi$-measurement an efficient alternative to knife-edge or post-selection slit-based experiments~\cite{Strekalov:1995ub,Pittman:1995jb,Howell:2004fc,AKjha2010,Just2013,Paul2014,Chen2019}. Here, we assume the apparent rotational symmetry of the JTMA in the joint $(q_x,q_y)$ planes, and scan the $\pi$-phase discontinuous profiles for both signal and idler SLMs along the $x$-axis.

\begin{figure*}[!ht]
\centering\includegraphics[width=1\textwidth]{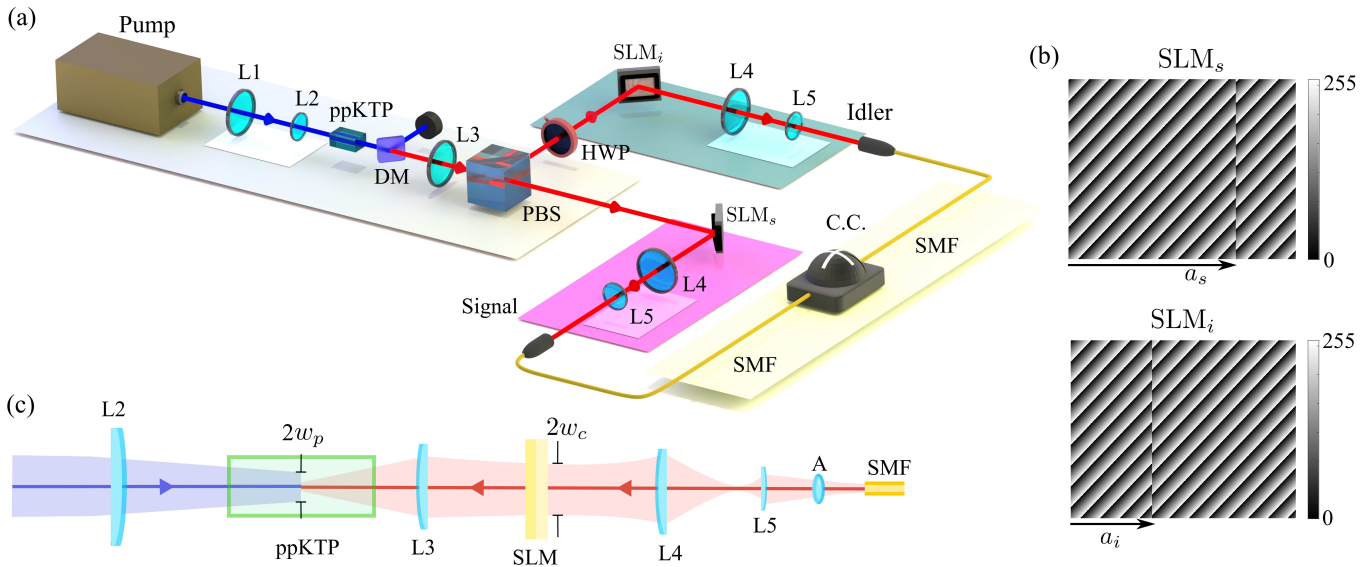}
\caption{\textbf{Experimental setup}: (a) A laser is used to pump a nonlinear ppKTP crystal to generate a pair of photons entangled in their transverse position-momentum via Type-II spontaneous parametric downconversion (SPDC). Our experiment is performed on two parallel setups with a continuous-wave laser diode pump at 405nm and a pulsed Ti:Sapph pump at 775nm. The pump photons are filtered by a dichroic mirror (DM) and the downconverted photons are separated with a polarising-beam-splitter (PBS).
The signal and idler photons are incident on two phase-only spatial light modulators (SLM) that are used for performing the $2D\pi$-measurement in the transverse position DoF.
The filtered photons are collected by a combination of telescopes (L4 and L5) and aspheric lenses (A) coupling to single-mode fibres (SMF), followed by detection through either single-photon-avalanche photodiodes (for $\lambda_{s1}, \lambda_{i1} = 810$nm) or  superconducting nanowire detectors (for $\lambda_{s2}, \lambda_{i2} = 1550$nm). A coincidence counting logic (CC) is used for recording time-coincident photon detection events within a coincidence window of 0.2~ns. (b) Examples of diffractive computer generated holograms implemented on the signal and idler SLMs for performing the $2D\pi$-measurement in the $x$-direction. (c) 2D profile of the experimental setup showing the relevant parameters $w_P$ and $w_C$. The pump beam radius ($w_P$) at the crystal plane determines the pump width parameter $\sigma_P$ (Eq.~\eqref{eq:SPDCparameters}). The back-propagated collection mode radius ($w_C$) at the SLM determines the collection width parameter ($\sigma_C$) at the crystal, which depends on the single-mode fibres and optical collection system lenses used.}
\label{fig:expsetup}
\end{figure*}

In particular, as illustrated in Fig.~\ref{fig:expsetup}b, the SLMs display
\begin{align}
    \Phi_n(q_{nx} ; a_n) =
    \begin{cases}
    1 & q_{nx} < a_n \\
    -1 & q_{nx} > a_n,
    \end{cases}
\end{align}
where $n\in\{s,i\}$ (signal/idler). By applying the hologram functions $\Phi_s$ and $\Phi_i$, the two-photon coincidence probability in Eq.~\eqref{eq:probcon} can be expanded to
\begin{align}
\begin{split}
\text{Pr}(a_s, a_i) &=\biggl| \int d q_{sy}d q_{iy}\bigg[\int^{a_s}_{-\infty} d q_{sx}\int^{a_i}_{-\infty} d q_{ix}G(\textbf{q}_s, \textbf{q}_i) \\ &+ \int_{a_s}^{\infty} d q_{sx}\int_{a_i}^{\infty} d q_{ix}G(\textbf{q}_s, \textbf{q}_i) - \int^{a_s}_{-\infty} d q_{sx}\int_{a_i}^{\infty} \\ & d q_{ix} G(\textbf{q}_s, \textbf{q}_i)  - \int_{a_s}^{\infty} d q_{sx}\int^{a_i}_{-\infty} d q_{ix}G(\textbf{q}_s, \textbf{q}_i)\bigg]\biggr|^2 
       \label{eq:twophotonprob}
\end{split}
\end{align}

Due to the Sinc dependence in Eq.~\eqref{eq:collectedJTMA}, the integrals in Eq.~\eqref{eq:twophotonprob} do not have any known closed analytic forms. Under certain approximations of $\sigma_P$ and $\sigma_S$, one can further simplify the expression in Eq.~\eqref{eq:twophotonprob} to ease its numerical evaluation and data fitting. For instance, the approximation of a nearly plane-wave pump beam ($\sigma_P \rightarrow 0$) simplifies Eq.~\eqref{eq:twophotonprob} to an integral that can be solved straightforwardly. However, the finite apertures in any optical system lead to a non-zero uncertainty in the momentum of the pump, thus resulting in an invalid approximation in practice.
\label{sec:exp}
Here, we discuss a more practical approximation that accounts for the effect of collection optics and is thus justified from an experimental point of view. We assert that when the generation bandwidth parameter $\sigma_S$ is sufficiently large compared to the collection bandwidth parameter $\sigma_C$ (whilst $\sigma_P$ remains relatively small), we can replace the Sinc argument in Eq.~\eqref{eq:collectedJTMA} with a Gaussian function. This relies on a comparison of the Gaussian envelopes determined by $\sigma_C$ and the Sinc envelope determined by $\sigma_S$.
In particular, transforming to the sum and difference coordinates, $\textbf{q}_\pm = \tfrac{1}{\sqrt{2}}(\textbf{q}_s\pm \textbf{q}_i)$ we have the collected JTMA
            \begin{align}
    \begin{split}
            \label{eq:collectedJTMASumDiff}
            &G(\textbf{q}_+, \textbf{q}_-) = \\ &\tfrac{\mathcal{N}_1}{\sqrt{\pi}\sigma_C}\exp \biggl\{\frac{-|\textbf{q}_+|^2}{\tilde{\sigma}_P^2}\biggr\} \exp\biggl\{\frac{-|\textbf{q}_-|^2}{\sigma_C^2}\biggr\}\textrm{sinc}\biggl(\frac{-2|\textbf{q}_-|^2}{\sigma_S^2}\biggr),
                \end{split}
\end{align}
where $\tilde{\sigma}_P = \bigl(\frac{1}{ \sigma_C^2}+\frac{1}{\sigma_P^2}\bigr)^{-\frac{1}{2}}$ and for $\sigma_P$ relatively smaller than $\sigma_C$, $\tilde{\sigma}_P \approx \sigma_P$. We then approximate the product of the Gaussian envelope ($\sigma_C$) and the Sinc argument ($\sigma_S$) to a Gaussian, which is valid if $\sigma_S \gtrapprox \sqrt{2}\sigma_C$ (see Appendix \ref{sec:Collected_to_CL}). Under this \textit{collection-limited} (CL) approximation, the \textit{collected} bi-photon JTMA ($G^{CL}$) reads
\begin{align}
\begin{split}
\label{eq:CLapprox}
    G(\textbf{q}_+, \textbf{q}_-) &\approx G^{CL}(\textbf{q}_+, \textbf{q}_-) \\ 
    &=\tfrac{\mathcal{N}_1}{\sqrt{\pi}\sigma_C}\exp \biggl\{-\frac{|\textbf{q}_s+\textbf{q}_i|^2}{2\tilde{\sigma}_P^2}\biggr\} \exp\biggl\{-\frac{|\textbf{q}_s-\textbf{q}_i|^2}{2\sigma_C^2}\biggr\}.
\end{split}
\end{align}
Notice that instead of $\sigma_S$, $\sigma_C$ now determines the width of the JTMA (see Fig.~\ref{fig:Coll_JTMA}b). Even though we have gotten rid of the Sinc dependence in the collection limited approximation, there is no analytic expression for the $2D\pi$-measurement. Still, we can derive some contributions analytically, from which we can recover $\sigma_P$ and $\sigma_C$. One of the relevant contributions is $\text{Pr}(a,-a)$, which is given as
\begin{align}
\text{Pr}(a,-a) = \bigg|N - N'\exp\bigg(-\frac{2a^2}{\sigma_C^2}\bigg)\biggr|^2,\label{eq:pxmxclosed}
\end{align}
where $N$ and $N'$ has absorbed all the constants that are independent of $a$. The expression of $\text{Pr}(a,-a)$ provides the information about the collection width parameter $\sigma_C$. Another crucial contribution that retrieves $\sigma_P$ is $\text{Pr}(a,a)$, which is given as
 \begin{align}
    \begin{split}
    \text{Pr}(a,a)  &=  \mathcal{A}\biggl[ \sqrt{2} \sigma_P\text{ } e^{-2a^2\big(\frac{1}{\tilde{\sigma}_P^2} + \frac{1}{\sigma_C^2}\big)} - 2\sqrt{\pi}\text{ }|a|\text{ } e^{\frac{-2a^2}{\sigma_C^2}} \\ &\times\text{ erfc}\bigg(\frac{\sqrt{2}|a|}{\tilde{\sigma}_P}\bigg) -\frac{\pi\sigma_C}{2\sqrt{2}}\bigg\{1-2\text{erf}\bigg(\frac{\sqrt{2}|a|}{\sigma_C}\bigg)\bigg\} \biggr]^2,\label{eq:pxx1}
    \end{split} 
\end{align}
where erfc is the complementary error function. 
By fitting the expressions Eq.~\eqref{eq:pxmxclosed} and Eq.~\eqref{eq:pxx1} to the experimental data corresponding to $\text{Pr}(a,-a)$ and $\text{Pr}(a,a)$, we can obtain $\sigma_C$ and $\sigma_P$ that describe the collected JTMA, which subsequently characterizes the spatial correlations produced in spontaneous parametric downconversion. For a detailed calculation of $\text{Pr}(a,-a)$ and $\text{Pr}(a,a)$, please refer to Appendix \ref{sec:theorycal}.

\section{Experiment and Results}
\begin{figure*}[ht]
\centering\includegraphics[width=\textwidth]{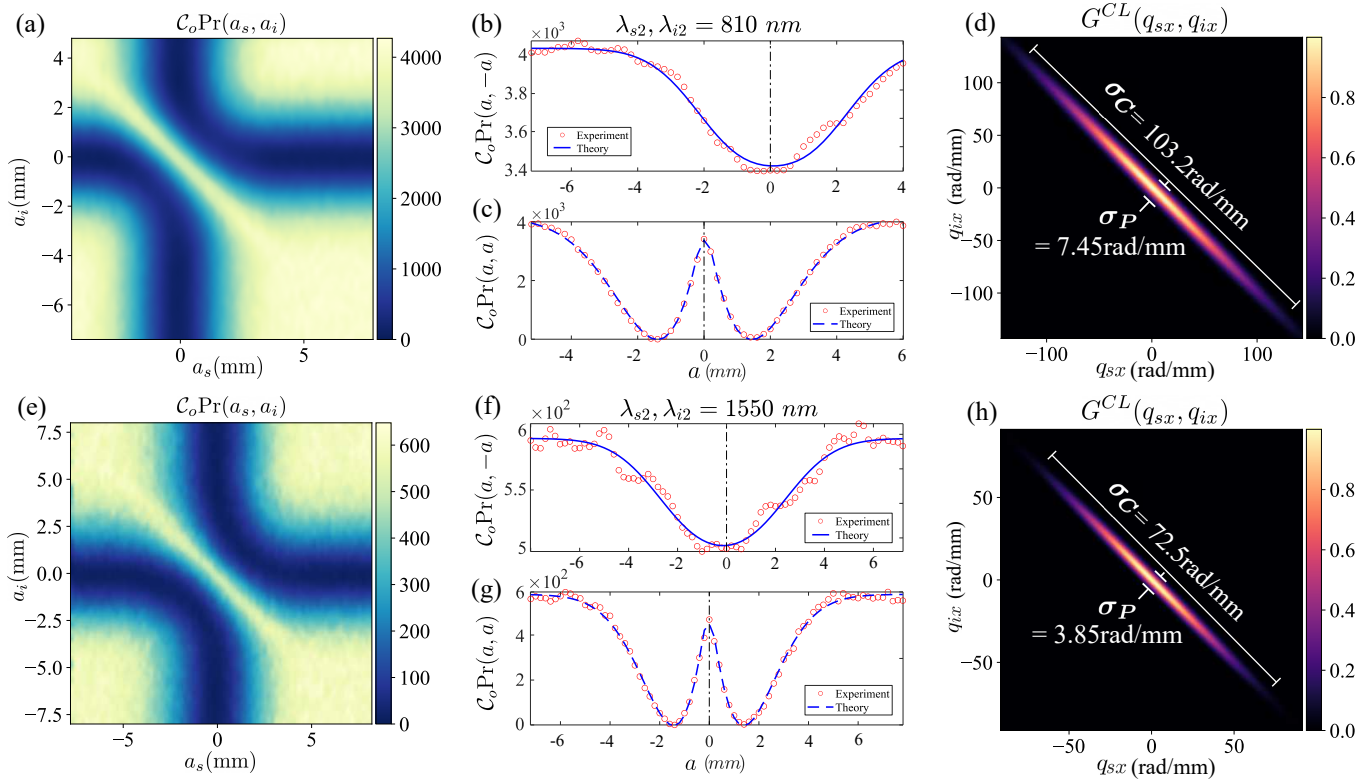}
\caption{\textbf{Experimental results}: $2D\pi$-measurement [$\mathcal{C}_o$Pr$(a_s, a_i)$] for (a) $\lambda_{s1},\lambda_{i1} = 810$nm and (e) $\lambda_{s2},\lambda_{i2} = 1550$nm, where $\mathcal{C}_o$ is proportional to two-photon coincidence count rates. We obtain $\sigma_C$ and $\sigma_P$ by fitting $\text{Pr}(a,-a)$ [(b) and (f)] and $\text{Pr}(a,a)$ [(c) and (g)]. Figures (d) and (h) show the experimentally determined collected JTMA at both wavelengths, where (d) for $\lambda_{s1},\lambda_{i1}=810$nm, $\sigma_C = 103.2\pm1.8$rad/mm and $\sigma_P = 7.45\pm1.51$rad/mm, and (h) for $\lambda_{s2},\lambda_{i2}=1550$nm, $\sigma_C = 72.5\pm2.3$rad/mm and $\sigma_P =3.85\pm0.31$rad/mm.}
\label{fig:exptheory}
\end{figure*}
To demonstrate the versatility of the $2D\pi$-measurement, we perform an experimental implementation at two different pump wavelengths (CW 405nm and femtosecond-pulsed 775nm).
The general setup is the same for both wavelengths (see Fig.~\ref{fig:expsetup}a). A laser is shaped by a telescope (L1 and L2) to pump a 5mm long periodically poled nonlinear ppKTP crystal that generates a pair of down-converted photons (for $405\text{nm}\rightarrow\lambda_{s1},\lambda_{i1} = 810$nm, for $775\text{nm}\rightarrow\lambda_{s2},\lambda_{i2} = 1550$nm)  entangled in their transverse position-momentum DoF via Type-II SPDC. 
After removing the pump with a dichroic-mirror (DM), the generated photons are separated with a polarising-beam-splitter (PBS) and made incident onto two phase-only spatial light modulators (SLM$_s$ and SLM$_i$) placed in the Fourier plane of the crystal via lens L3 ($f = 250$mm). 
The spatial field at the SLM plane is directly related to the transverse momentum space at the crystal plane via $\textbf{q}=\tfrac{2\pi x}{f \lambda}$, where $f$ is the focal length of L3 and $\lambda$ is the signal/idler wavelength. 
We perform the $2D\pi$-measurement using diffractive holograms displayed on the SLMs (Fig.~\ref{fig:expsetup}b), together with the collection of single-mode fibres (SMFs), allow for arbitrary spatial-mode projective measurements to be performed on the incoming photons. 

The optical system parameters (lenses L1--L5 and A) are judiciously chosen. First, in order to obtain a highly correlated JTMA, the telescope system of lenses L1 and L2 is chosen to maximise the pump radius $w_P$ at the crystal plane, thus minimising the pump width parameter $\sigma_P$ for the strength of the momentum correlation, while ensuring that the pump beam is not truncated by the crystal aperture. Next, consider the back-propagated beam from the SMF to the ppKTP crystal (shown in red in Fig.~\ref{fig:expsetup}c). The aspheric lens A and the optical system of lenses L3--L5 are chosen such that the collection width parameter $\sigma_C$ meets the condition $\sigma_S\gtrapprox \sqrt{2} \sigma_C$, allowing us to work under the collection-limited JTMA approximation (see previous section). 
The telescope system L4 and L5 has also been referred to in our previous work as an ``intensity-flattening telescope'' (IFT) as it effectively broadens the back-propagated Gaussian envelope of the collection mode such that higher order modes associated with the edges of the JTMA are measured efficiently, while the lower order modes are suppressed \cite{Bouchard:2018hr}.

The photons are detected by single-photon-avalanche photodiodes for $\lambda_{s1},\lambda_{i1} = 810$nm and superconducting nanowire detectors (SNSPD) for $\lambda_{s2},\lambda_{i2} = 1550$nm, which are connected to a coincidence counting logic (CC) with a coincidence window of 0.2ns. We characterize the collected JTMA at the Fourier plane of the crystal located at the SLM planes. The plots in Figs.~\ref{fig:exptheory}a and \ref{fig:exptheory}e show the data obtained for the $2D\pi$-measurement performed at both wavelengths ($\lambda_{s1},\lambda_{i1} = 810$nm and $\lambda_{s2},\lambda_{i2} = 1550$nm), while the reconstructed JTMAs are shown in Figs.~\ref{fig:exptheory}d and \ref{fig:exptheory}h. We obtain $\sigma_C$  and $\sigma_P$ by fitting the closed-form expression of $\text{Pr}(a,-a)$ and $\text{Pr}(a,a)$ (Eqs.~\eqref{eq:pxmxclosed} and \eqref{eq:pxx1}) to the experimental data. 
It is worth noting that the feature corresponding to $\sigma_P$ is also present in the visibility of $\text{Pr}(a,-a)$, which is shown in the fitting curves in Figs.~\ref{fig:exptheory}b and ~\ref{fig:exptheory}f (refer to Eq.~\eqref{eq:visbilityinpraa}). Those features, therefore, provide a sensitive measure of the correlation strength $\sigma_P$ even when the resolution of the scan is coarse, unlike slit-based measurements where the trade-off between the slit size and photon flux is the issue.

\begin{bigboxtable}[floatplacement=t,halign = center]{Predicted and measured parameters describing the generated and collected joint-transverse-momentum-amplitude (JTMA) obtained from the $2D\pi$-measurement.}{expvalues}
\begin{center}
\begin{tabularx}{\textwidth}{|p{0.1\textwidth}|p{0.13\textwidth}|p{0.15\textwidth}p{0.15\textwidth}|p{0.13\textwidth}|p{0.134\textwidth}p{0.14\textwidth}|}
\hline 
$\lambda$ & ${w_p}$ & $\sigma_P^{\textrm{pre}}$ & $\sigma_P^{\textrm{meas}}$ & $\sigma_S^{\textrm{pre}}$& $\sigma_C^{\textrm{pre}}$& $\sigma_C^{\textrm{meas}}$\\
(nm) &($\mu$m) & (rad/mm) & (rad/mm) & (rad/mm) & (rad/mm) & (rad/mm)\\
\hline 
810 & $188\pm 5.3$ & $7.52\pm 0.21$ & $7.45\pm 1.51$ & $151.1 \pm 3.1$ & $106.5 \pm 12.6$ & $103.2\pm 1.8$  \\
1550 & $450\pm 5.3$ & $3.14\pm0.04$ & $3.85\pm0.48$ & $106.7\pm 2.1$ & $76.7\pm 8.01$& $72.5\pm2.3$\\
\hline
\end{tabularx}
\end{center}
\tcblower
\footnotesize{The predicted pump width and generation width parameters ($\sigma_P^{\textrm{pre}}$ and $\sigma_S^{\textrm{pre}}$) are obtained from the pump waist $w_p$ and crystal length according to Eq.~\eqref{eq:SPDCparameters}. The predicted collection width parameter ($\sigma_P^{\textrm{pre}}$) is calculated from measurements of the optical collection system (please see main text for details). The measured parameters ($\sigma_P^{\textrm{meas}}$ and $\sigma_C^{\textrm{meas}}$) are obtained from the $2D\pi$-measurement performed on the two-photon state. Both predicted and measured values of $\sigma_C$ meet the condition $\sigma_S \gtrapprox \sqrt{2}\sigma_C$ that allows us to operate under the collection-limited (CL) approximation (please see Eq.~\eqref{eq:CLapprox}).}
\end{bigboxtable}
The measured values of the pump and collection width parameters ($\sigma_P^{\textrm{meas}}$, $\sigma_C^{\textrm{meas}}$) obtained from the $2D\pi$-measurement are reported in Table~\ref{table:expvalues} and agree with their predicted values ($\sigma_P^{\textrm{pre}}$, $\sigma_C^{\textrm{pre}}$), which are calculated from our knowledge of the optical system parameters. The predicted value of the pump width parameter $\sigma_P^{\textrm{pre}}$ for both wavelengths is calculated from the $1/e^2$ pump radius at the crystal plane (Eq.~\eqref{eq:SPDCparameters}), and the predicted collection width parameter $\sigma_C^{\textrm{pre}}$ is calculated by back-propagating the width of the fundamental Gaussian mode of the SMFs to the crystal plane (Fig.~\ref{fig:expsetup}c). The error propagation is analysed by taking into account $\pm$0.5mm uncertainties of the measured distances between lenses and focal lengths.

\section{Tailoring High-Dimensional Entanglement}
\begin{figure*}[t]
\centering\includegraphics[width=\textwidth]{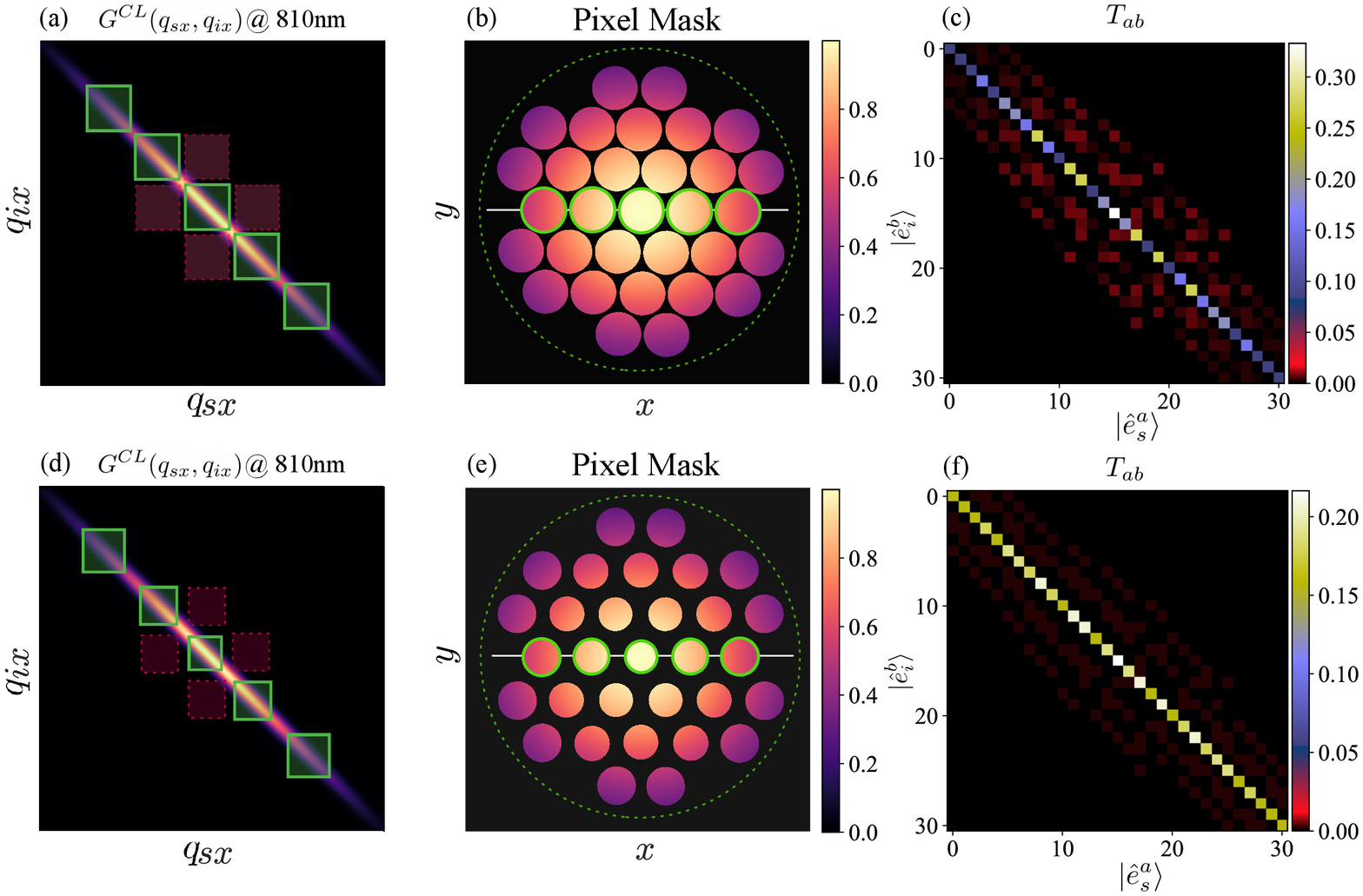}
\caption{\textbf{Optimising $T_{ab}$ for 810nm two-photon state by tailoring the pixel mask in 31 dimensions:} Figures (a-c) [(d-f)] show the collected JTMA, pixel mask holograms, and resulting state coefficients $T_{ab}$, before [after] optimisation. The square regions in the JTMA plots in (a,d) indicate the values of $q_{sx}$ and $q_{ix}$ used to generate the pixel masks holograms in (b,e). The green square regions represent the pixels that are correlated ($T_{aa}$ elements), while the red ones are responsible for cross-talk between them ($T_{ab}$ elements $\forall a\neq b$). Note that while the 2D JTMA ($x$-dimension) is shown here, calculating all the elements of $T_{ab}$ require integration over the entire 4D expression for the JTMA (Eq.~\ref{eqs:tabElements}), i.e., $q_{sx}$, $q_{sy}$, $q_{ix}$ and $q_{iy}$. The green dashed circles represent the maximum size of the detectable area, which is further constrained by the finite collection bandwidth parameter $\sigma_C$. In (a) and (b), the pixels are set to the same size with a fixed spacing between them, and the resulting state (c) shows significant cross-talk between discrete spatial modes. The dimensionality of entanglement ($d_{ent}$) is calculated to be 26 in the Hilbert space of $d =31$ pixels~\cite{Bavaresco:2018gw}. In (d) and (e), we show how one can tailor the pixel mask from knowledge of the JTMA such that the coincidence probability $\text{Pr}(\Phi_s,\Phi_i)$ is equal for all pixels, while minimizing the cross-talk between them at the expense of total counts. This  ensures that the state shown in (f) is optimized to be in the Schmidt basis while also being close to maximally entangled. Consequently, the dimensionality of entanglement is increased to its maximal value of $d_{ent} = 31$. }
\label{fig:Tab_before_after}
\end{figure*}

Once accurate knowledge of the continuous position-momentum two-photon state characterised by the JTMA has been obtained, one may want to discretise such a state for use in quantum information applications based on discrete variables \cite{Dada:2011dn,Mafu:2013jk,Bavaresco:2017if,HerreraValencia2020highdimensional}. For instance, to harness discrete variable high-dimensional entanglement, one needs to choose an appropriate modal basis in which to work.
The design of such discrete modal bases is often informed by the Schmidt decomposition of the entire bi-photon wavefunction~\cite{Straupe:2011ju,Walborn2012}. However, at the expense of lower count rates one can design modal bases to optimise for more general figures of merit such as entanglement-of-formation ($E_{\textrm{oF}}$), heralding efficiency, and measurement fidelity, while taking into account the types of devices used (for example, phase-only SLMs). 

We begin with a theoretical treatment of how our continuous two-photon state is discretised via specific projective modal measurements. In practice, we display holograms $\{\Phi_s(\textbf{q}_s)\}^a$ and $\{\Phi_i(\textbf{q}_i)\}^b$ (where $|\Phi_s(\textbf{q}_s)|\leq1$) on the SLMs to generate a subnormalised post-selected state $\ket{\psi^{(PS)}}$ in the \emph{standard} discrete modal basis $\{\ket{\hat{e}_n}\}_a$:
\begin{equation}
    \ket{\psi^{(PS)}} =  \sum_{ab} T_{ab}  \ket{ \hat e_{s}^a} \ket{ \hat e_{i}^b},
    \label{eq:postselected}
\end{equation}
where the complex elements $T_{ab}$ are given by
\begin{equation}
\label{eqs:tabElements}
    T_{ab} =\int d^2\textbf{q}_{s}\int d^2\textbf{q}_{i} \Phi_{s}^a(\textbf{q}_{s}) \Phi_{i}^b(\textbf{q}_{i}) G(\textbf{q}_{s},\textbf{q}_{i}).
\end{equation}
Here, $G(\textbf{q}_{s},\textbf{q}_{i})$ is the collected JTMA (collection-limited in our case) whose form can be obtained via the 2D$\pi$ measurement described in the preceding sections. The holograms $\Phi_{s}^a(\textbf{q}_{s})$ and $\Phi_{i}^b(\textbf{q}_{i})$ can be constructed in a manner to ensures that the associated discrete modal bases are orthonormal (see Appendix \ref{sec:DesginApp}). 

Now we can determine the probabilities associated with measuring generalised projectors (arbitrary coherent superpositions) both in the two-photon and single-photon case. We can measure an arbitrary normalised vector $\ket{\vec{v}_s} = \sum_a v^a_s \ket{\hat{e}^a_s}$ by constructing hologram $\Phi_s^{\vec{v}}(\textbf{q}_s)$ given by
\begin{equation}
    \Phi^{\vec v_s}_s(\textbf q_s) = A^{\vec v_s} \sum_a v_s^a \Phi^a_s(\textbf q_s),
\end{equation}
with $A^{\vec v}$ chosen such that $\max_{\textbf q} |\Phi^{\vec v}_s(\textbf q)| \leq 1$. The preceding condition ensures that a hologram does not increase energy and can only add loss. The two-photon coincidence probability for measuring our state in modes $\ket{\vec{v}_s}$ and $\ket{\vec{v}_i}$ is given by
\begin{equation}
    \textbf{Pr}(\vec v_s, \vec v_i) = (A^{\vec v_s}A^{\vec v_i})^2 |\bra{\vec v_s}\bra{\vec v_i}\ket {\psi^{(PS)}}|^2.
\end{equation}\label{eq:coincprobeff}
Similarly, the probability of measuring a signal photon (inclusive of a possible idler photon) in mode $\ket{\vec{v}_s}$ depends on the collection mode of only signal $(\mathcal{C}(\textbf q_s|\sigma_C))$ and is expressed as
\begin{equation}
\begin{split}
\label{eq:singlePhotonStatisticsMain}
\textbf{Pr}(\vec{v}_s)&= \int d^2 \textbf{q}_i\left|\int d^2 \textbf{q}_s\Phi^{\vec v_s}_s(\textbf q_s)\mathcal{C}(\textbf q_s|\sigma_C) F(\textbf q_s,\textbf q_i) \right|^2,\\
\end{split}
\end{equation}
where the product of generated JTMA $F(\textbf q_s,\textbf q_i)$ and the signal's collection mode $\mathcal{C}(\textbf q_s|\sigma_C)$ can be further simplified as discussed in Appendix~\ref{sec:Collected_to_CL} (see Eq.~\eqref{eq:sincprodgauss}).

 With knowledge of the collected JTMA ($G(\textbf{q}_s,\textbf{q}_i)$), a variety of figures of merit might be considered when designing such modal bases (corresponding to holograms $\Phi_{s}^a(\textbf{q}_{s})$ and $\Phi_{i}^b(\textbf{q}_{i})$) and choosing which measurements to make in a given basis. Here, we take the example of disjoint discrete spatial modes (``pixel basis'' \cite{HerreraValencia2020highdimensional}) defined by macro-pixels in the SLM plane to exemplify some of these merits. In the pixel basis, the size of each pixel, the spacing between them, and the size of the complete pixel mask are important parameters to take into account in the basis design. One may vary these parameters to optimise for the following properties:\\
 
 \noindent\textbf{Schmidt basis:} A standard discrete basis for the post-selected two-photon state (Eq.~(\ref{eq:postselected})) can be designed such that it corresponds to the Schmidt basis where the coincidence cross-talk between modes is minimized, and thus suppressing the off-diagonal elements of $T_{ab}$ ($T_{ab}\rightarrow 0$ $\forall a\neq b$). In the case of the pixel basis, this corresponds to choosing the spacing between pixels to be at least equal to the pump width parameter (appropriately propagated to position coordinates at the SLM plane), which determines the JTMA correlation strength  $\sigma_P^{(x)}=\frac{f \lambda}{2 \pi}\sigma_P$ (see Fig.~\ref{fig:Tab_before_after}).\\
 
 \noindent\textbf{Entanglement dimensionality:} When constructing a standard discrete basis, there is a limit on the entanglement dimensionality---the maximal number of correlated modes that can be considered whilst remaining in the Schmidt basis. Information about this can be deduced from the JTMA: the accessible number of \textit{generated} entangled modes, related to the reciprocal of the marginal state purity (often known as the Schmidt number~\cite{Law:2004hw}), can be estimated through $\sigma_P$ and $\sigma_S$~\cite{Schneeloch:2016ch}. However, as we have shown here, $\sigma_S$ is often constrained by the collection width parameter $\sigma_C$, knowledge of which can be used to estimate the perhaps more relevant number of \textit{collected} entangled modes. For the pixel basis, this involves an optimisation of the number of correlated macro-pixels one can fit within the collected area, while having appreciable count rates.\\ 
 
\noindent\textbf{Maximal entanglement:} Optimising the standard discrete basis such that the coincidence probability for each mode is equal (while remaining in the Schmidt basis) imposes $T_{ab}$ is proportional to the identity matrix and thus a maximally entangled state. This maximises entropic quantifiers of entanglement such as entanglement-of-formation ($E_{\textrm{oF}}$) and can be achieved for instance, by optimally varying pixel size as a function of radial distance from the optic axis (see Fig.~\ref{fig:Tab_before_after} for a detailed example).\\
    
\noindent\textbf{Basis-dependent efficiency:} One can find bases where all the holograms can be efficiently realised by maximising $A^{\vec v}$ for all elements of a basis. For instance, with disjoint pixels, all bases mutually unbiased to the standard basis obtain $A^{\vec v}=\sqrt{d}$ for all elements. This ensures that all measurements maximise photon flux, thus drastically reducing measurement times~\cite{HerreraValencia2020highdimensional}.\\

\noindent\textbf{Heralding efficiency:} The heralding efficiency, or the probability that the detection of a photon in one mode (signal) indicates a photon in the other (the heralded photon or idler), is normally studied in a symmetric configuration, i.e.~the same collection parameters apply to both photons~\cite{PhysRevA.83.023810,PhysRevA.90.043804}. The inherent multi-mode nature of the JTMA opens up an alternate way to tune heralding efficiencies in an asymmetric manner, i.e.~with different collection parameters and resultant heralding efficiencies for each photon~\cite{PhysRevA.72.062301}. We can define a one-sided heralding efficiency in this case, where measuring the signal photon in mode $\ket{\vec{v}_s}$ heralds the presence of an idler photon in mode $\ket{\vec{v}_i}$ with an efficiency
\begin{equation}
    \eta^{s\rightarrow i} = \frac{\textbf{Pr}(\vec v_s, \vec v_i)}{\textbf{Pr}(\vec{v}_s)}. 
    \label{eq:heraldingEffi}
\end{equation}
Therefore, designing holograms $\Phi_i^{\vec{v}}(\textbf{q}_i)$ to maximise Eq.~(\ref{eq:heraldingEffi}), and choosing bases such that $A^{\vec v_i}$ is large can lead to high one-sided heralding efficiency (see Appendix \ref{sec:DesginApp}). 
For instance, increasing the size of the idler pixel than that of the signal optimises the heralding efficiency, resulting in a larger overlap of heralded photons on the idler side.\\

We have recently implemented some of the above techniques experimentally in order to rapidly certify high-fidelity entangled states with entanglement dimensionalities up to $d=55$, entanglement-of-formation up to $E_{\textrm{oF}}=4$ ebits~\cite{HerreraValencia2020highdimensional}, and to violate high-dimensional steering inequalities in dimension up to $d=15$ \cite{designolle2020genuine}. These experiments were performed in the pixel basis, where through knowledge of the JTMA obtained via the $2D\pi$-measurement, a post-selected state closest to a maximally entangled state was realised. The SLM holograms were designed to minimise the cross-talk between modes, while simultaneously equalising photon count rates for all discrete modes across $\sigma_C$ (in a manner similar to the procedure discussed in Fig.~\ref{fig:Tab_before_after}). Furthermore, high dimensional entanglement witnesses exploiting only bases mutually unbiased to the standard pixel basis ensured high basis-dependent efficiencies and minimised measurement times. 
\begin{figure}[t!]
\centering\includegraphics[width=1\columnwidth]{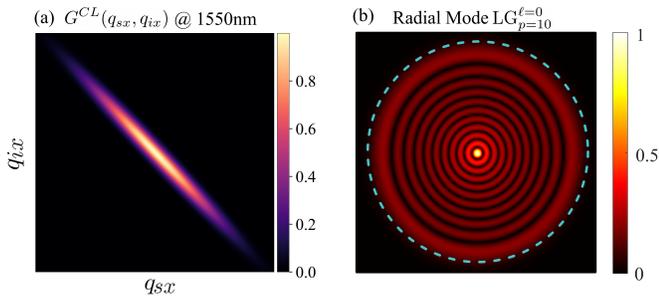}
\caption{\textbf{LG radial-mode entanglement}: (a) The knowledge of collected JTMA ($G(\textbf{q}_s,\textbf{q}_i)$) determines the maximum size of the mode waist (blue dotted circle in (b)) to witness entanglement in 7 dimensions in $p=0,\dots,10$ radial LG modes~\cite{valencia2021entangled}.}
\label{fig:LG_jtma}
\end{figure}

While the above examples have used the disjoint and discrete pixel basis, knowledge of the JTMA can be used to optimise other spatial mode bases such as the Laguerre-Gaussian (LG) basis, which plays a significant role in classical and quantum optics~\cite{Krenn2017}. In the LG basis, correlations in both the azimuthal and radial components depend on the relationship between the pump and the down-converted signal/idler mode waists~\cite{Miatto:2011cr,Salakhutdinov2012}, indicating the effective number of detectable Schmidt modes that are entangled in the full transverse field~\cite{Miatto:2012bv,Straupe:2011ju}. The ability to determine the collected JTMA allows us to experimentally adjust $\sigma_P$ and $\sigma_C$ such that the correlations are maximised. While the degree of correlations relates to $\sigma_P$, information of $\sigma_C$ sets a limit on the size of modes we can optimally measure (see Fig.~\ref{fig:LG_jtma}), which is of particular importance when dealing with modes that have radial dependence. Taking these considerations into account, we were able to recently certify entanglement dimensionalities of up to 26 in a 43-dimensional radial and azimuthal LG space~\cite{valencia2021entangled}, demonstrating the potential of the JTMA in harnessing the full capabilities of high-dimensional entanglement.

\section{Conclusion and Discussion}
In our work, we have studied the spatial wavefunction of position-momentum entangled bi-photon states generated in collinear Type-II SPDC. We define the collected joint-transverse-momentum amplitude (JTMA), a function that characterizes bi-photon state in the momentum degree-of-freedom while incorporating the effects of the measurement system. 
We propose a method to efficiently and accurately characterize the collected JTMA using phase-only modulated holograms, and experimentally demonstrate it on two identical entanglement sources at different wavelengths.
From knowledge of the collected JTMA, we discuss how one can tailor discrete-variable high-dimensional entangled states via projective measurements in several spatial mode bases. 
Our techniques can be used to generate high-fidelity, high-dimensional entangled states of light, which can be further optimised for properties such as maximal entanglement and single photon heralding efficiencies.
The utility of our characterisation methods is evident in some recent works, where we have used knowledge of the JTMA to tailor diverse kinds of high-dimensional entangled states of light with record quality and dimensionalities in device-dependent~\cite{HerreraValencia2020highdimensional, valencia2021entangled} as well as one-sided device-independent platforms~\cite{designolle2020genuine}.

Information about the JTMA could be used for implementing and optimising arbitrary spatial-mode projective measurements, which are required for violating Bell-like inequalities proposed for high-dimensional systems~\cite{Vertesi:2010bq,CGLMP,SATWAP}. Additionally, knowledge of the JTMA can be used for tuning one-sided photon heralding efficiencies~\cite{Iskhakov:2015va, Ramelow:2013dz, Castelletto:2005wt}, which play a significant role in device-independent tests of quantum mechanics and the related field of device-independent QKD. 
The ability to control the correlations/anti-correlations of an entangled pair of photons using the measured JTMA parameters could enable the engineering of quantum states with tailored spatial and spectral properties~\cite{Brambilla2010}, which could be used to boost the performance of quantum-enhanced imaging and metrology~\cite{Malik:2014ht,Basset:2019cw}. In addition, our methods for characterizing the JTMA can be translated to other degrees-of-freedom such as time-frequency \cite{Jha:2008,Kues:2017,Martin:2017}, and could enable the characterization of the full spatio-temporal bi-photon wavefunction, which would have a wide-ranging impact on entangled-based quantum technologies. 

 \begin{acknowledgements}
  This work was made possible by financial support from the QuantERA ERA-NET Co-fund (FWF Project I3773-N36), the UK Engineering and Physical Sciences Research Council (EPSRC) (EP/P024114/1), and the European Research Council (Starting Grant PIQUaNT).
 \end{acknowledgements}
\bibliographystyle{apsrev4-1fixed_with_article_titles_full_names}
\bibliography{references}
\clearpage
\onecolumngrid
\appendix

\section{Joint-transverse-momentum-amplitude (JTMA) of Type-II SPDC with Periodic Poling}\label{sec:JTMAapp}
We consider the case of Type-II SPDC (whereby the pump, signal and idler are polarised on the $e$, $e$ and $o$ axes respectively), for periodically poled crystals designed to achieve phase-matching at degenerate signal/idler frequencies in the colinear configuration.
To account for all these contributions, we derive an expression for the entire bi-photon state, and apply approximations to arrive at the JTMA stated in the main text (Eq.~\eqref{eq:JTMAformula}). Following the conventional asymptotic fields approach, the first-order nonlinear Hamiltonian in the backward-Heisenberg picture, $H^{NL}(t)$ (see reference~\cite{Yang2008} for a detailed account) takes the form,
\begin{equation}
\begin{split}
        \int H^{NL}(t) dt \propto \int dt\int d^3\textbf{k}_i \int d^3\textbf{k}_s \int d^3\textbf{k}_p \int d^3 \textbf{x} \exp{(\textit{i}(\textbf{k}_s+ \textbf{k}_i -\textbf{k}_p)\cdot\textbf{x}})\chi^{(2)}(\textbf{x})  \\ \times \exp{(\textit{i}(\omega(\textbf{k}_s) +\omega(\textbf{k}_i) - \omega(\textbf{k}_p))t)}\text{ }\alpha_p(\textbf{k}_p) \hat{a}_s^{\dagger}(\textbf{k}_s) \hat{a}_i^{\dagger}(\textbf{k}_i) + h.c.
\end{split}\label{eq:HNL}
\end{equation}
The temporal integral can be extended over all time, leading to the energy-matching term, $\int dt \exp{(\textit{i}\Delta\omega t)} = \delta(\Delta \omega) = \delta( \omega(\textbf{k}_s) +\omega(\textbf{k}_i) -\omega(\textbf{k}_p))$. 
The $d^3 \textbf{x}$ integral over a crystal of dimensions $L_x^\perp \times L_y^\perp \times L_z$, situated about the origin can be evaluated as, 
\begin{equation}
 \begin{split}
\int_{-L_x^\perp/2}^{L_x^\perp/2}&dx_x\int_{-L_y^\perp/2}^{L_y^\perp/2}dx_y\int_{-L_z/2}^{L_z/2} dx_z \exp{(\textit{i}(\textbf{k}_s+ \textbf{k}_i -\textbf{k}_p)\cdot\textbf{x}})\chi^{(2)}(\textbf{x})\\
&=\sinc\left(\frac{L_x^\perp}{2}(q_{sx}+q_{ix}-q_{px})\right)\sinc\left(\frac{L_y^\perp}{2}(q_{sy}+q_{iy}-q_{py})\right)\sinc\left(\frac{L_z}{2} \Delta k_z \right)\\
&\approx \delta(\textbf{q}_s+\textbf{q}_i-\textbf{q}_p)\sinc\left(\frac{L_z}{2} \Delta k_z \right),
 \end{split}\label{eq:xint}
\end{equation}
where we introduce notation $\textbf{q}_n$ for the transverse components of $\textbf{k}_n$ (for $n=s,i,p$), and the longitudinal wavevector mismatch, $\Delta k_z$ (to be defined), contains a contribution from the periodic poling structure with the period $\Lambda$. The final approximation will hold for crystals with transverse extents ($L_x^\perp$, $ L_y^\perp$) that are sufficiently large.

Using the approximations above, the integral in Eq.~\eqref{eq:HNL} can be written as
\begin{equation}
\begin{split}
        \int H^{NL}(t) dt \propto \int d^3\textbf{k}_i \int d^3\textbf{k}_s \int d^3\textbf{k}_p \delta(\Delta \omega) \delta(\textbf{q}_s+\textbf{q}_i-\textbf{q}_p)\sinc\left(\frac{L_z}{2} \Delta k_z \right) \alpha_p(\textbf{k}_p) \hat{a}_s^{\dagger}(\textbf{k}_s) \hat{a}_i^{\dagger}(\textbf{k}_i) + h.c.
\end{split}\label{eq:HNL2}
\end{equation}

\noindent The frequency of each field is dependent only on the modulus of the wave-vector (and its polarisation, assuming close to colinear propagation). Furthermore, we need to consider fields only with positive $z$ components, which invites a change of integration variables from Cartesian components of $\textbf{k}$ to transverse wave-vector, $\textbf{q}$, and the modulus of the wave-vector, $|\textbf{k}|$. Note that such a coordinate change results in a lack of clear correspondence of the mode operators to their strictly orthonormal counterparts, however this mathematical convenience remains suitable for obtaining a description of the biphoton state.  
Hence, we can express the $z$-components of $\textbf{k}$'s in terms of their modulus and transverse components, 
\begin{equation}
    k_z = \sqrt{\abs{\textbf{k}}^2 - \abs{\textbf{q}}^2} = |\textbf{k}|\sqrt{1 - \frac{|\textbf{q}|^2}{|\textbf{k}|^2}} \simeq |\textbf{k}|\bigg(1 - \frac{1}{2}\frac{|\textbf{q}|^2}{|\textbf{k}|^2} \\ -\frac{1}{8}\frac{|\textbf{q}|^4}{|\textbf{k}|^4} +O(\frac{|\textbf{q}|^6}{|\textbf{k}|^6})\bigg),
\end{equation}
where we make use of a close-to-colinear approximation, $|\textbf{q}|\ll|\textbf{k}|$. Differentiating w.r.t.~$|\textbf{k}|$, we have
\begin{equation}
        d k_{z} = (1 - \frac{|\textbf{q}|^2}{|\textbf{k}|^2})^{-1/2} d|\textbf{k}| \simeq (1 +\frac{1}{2}\frac{|\textbf{q}|^2}{|\textbf{k}|^2} +  O(\frac{|\textbf{q}|^4}{|\textbf{k}|^4}))d|\textbf{k}|,\label{eq:normcorr}
\end{equation}
allowing the transformation,
\begin{equation}
    d^3 \textbf{k} = d^2\textbf{q} \, d k_z
    \approx d^2\textbf{q} \,
    d|\textbf{k}|\,(1 +\frac{1}{2}\frac{|\textbf{q}|^2}{|\textbf{k}|^2}).\label{eq:kapprox} 
\end{equation}

\noindent The phase-matching contribution arising from the longitudinal wave-vector mismatch can now be expressed as
\begin{equation}
 \begin{split}
     \Delta k_z &= k_{sz} +k_{iz} -k_{pz}+\tfrac{2\pi}{\Lambda} \\
    &= \sqrt{|\textbf{k}_s|^2 - |\textbf{q}_s|^2} + \sqrt{|\textbf{k}_i|^2 - |\textbf{q}_i|^2} - \sqrt{|\textbf{k}_p|^2 - |\textbf{q}_p|^2}+\tfrac{2\pi}{\Lambda} \\
    &= |\textbf{k}_s|\sqrt{1 - \frac{|\textbf{q}_s|^2}{ |\textbf{k}_s|^2}} + |\textbf{k}_i|\sqrt{1 - \frac{|\textbf{q}_i|^2}{ |\textbf{k}_i|^2}} - |\textbf{k}_p|\sqrt{1 - \frac{|\textbf{q}_p|^2}{ |\textbf{k}_p|^2}} +\tfrac{2\pi}{\Lambda}\\
    &\approx |\textbf{k}_s| + |\textbf{k}_i| - |\textbf{k}_p| -\frac{1}{2}\bigg[\frac{|\textbf{q}_s|^2}{ |\textbf{k}_s|} +\frac{|\textbf{q}_i|^2}{ |\textbf{k}_i|} - \frac{|\textbf{q}_p|^2}{ |\textbf{k}_p|}\bigg]+\tfrac{2\pi}{\Lambda}
 \end{split}
\end{equation}
To proceed, we make use of the dispersion relations for the various fields, for which we make first-order expansions (no group velocity dispersion)  about the degenerate, energy-matched frequencies $\omega_0:=\omega_{s0}=\omega_{i0}=\omega_{p0}/2$, so that for the pump field,
\begin{equation}
\begin{split}
    |\textbf{k}_p|=k^{(e)}_p(\omega_p) \simeq k^{(e)}_p(\omega_{p0}) +(\omega_p - \omega_{p0})\frac{1}{v_{gp}} = k_{p0} +\frac{\tilde{\omega}_p}{v_{gp}}= k_{p0} +\tilde{k}_p
\end{split}    
\end{equation}
where $k^{(e)}_p(\omega_p)$ is the dispersion relation for polarisation on the extraordinary axis expanded about $\omega_{p0}$ (the central pump frequency), $\tilde{\omega}_p=(\omega_p - \omega_{p0})$ is the frequency offset, and $v_{gp}$ and $k_{p0}$ are the group velocity and wave-vector of the pump field at $\omega_{p0}$, polarised accordingly. Similarly, for signal and idler fields,
\begin{equation}
\begin{split}
    |\textbf{k}_s|=k^{(e)}_s(\omega_s) \simeq k^{(e)}_s(\omega_{0}) +(\omega_p - \omega_{0})\frac{1}{v_{gs}} = k_{s0} +\frac{\tilde{\omega}_s}{v_{gs}}= k_{s0} +\tilde{k}_s \\
        |\textbf{k}_i|=k^{(o)}_i(\omega_i) \simeq k^{(o)}_i(\omega_{0}) +(\omega_i - \omega_{0})\frac{1}{v_{gi}} = k_{i0} +\frac{\tilde{\omega}_i}{v_{gi}}= k_{i0} +\tilde{k}_i
\end{split}{}    
\end{equation}{}
\noindent This allows us to consider the wave-vector mismatch for colinear generation ($\textbf{q}\rightarrow 0$) at degeneracy ($\tilde \omega \rightarrow 0$),
\begin{equation}
\begin{split}
\Delta k_{z0}:=k_{s0}+k_{i0}-k_{p0}.
\end{split}   
\end{equation}
\noindent Thus, rewriting $\Delta k_z$ at degeneracy as
\begin{equation}
    \Delta k_z = \Delta k_{z0} + \tfrac{2\pi}{\Lambda}
\end{equation}
For sources optimised in this regime, periodic poling of the crystal is used to cancel this contribution, thereby achieving phase-matching ($\Delta k_z = 0$). Thus, we set the poling period, $\tfrac{2\pi}{\Lambda}=-\Delta k_{z0}$. 

Noting that we may write the energy-matching term as $ \delta(\Delta \omega)=v^{-1}_{gp} \delta(\tfrac{v_{gs}}{v_{gp}} \tilde k_s +\tfrac{v_{gi}}{v_{gp}} \tilde k_i - \tilde k_p)$, then imposing that the pump field may be expressed as a separable function of $\textbf{q}$ and $\tilde k$ so that $\alpha_p(\textbf{k})\approx \alpha_p(\textbf{q})\tilde \alpha_p(\tilde k)$, and using the approximation made in Eq.~\eqref{eq:kapprox}, we can perform the integrals $d^3\textbf{k}_p$ in Eq.~\eqref{eq:HNL2} to arrive at the form,
\begin{equation}
    \begin{split}
             \int H^{NL}(t) dt &\propto \int d^2\textbf{q}_i d \tilde k_i\int d^2\textbf{q}_s d \tilde k_s v^{-1}_{gp}\times\bigg(1 +\frac{1}{2}\frac{|\textbf{q}_s|^2}{(k_{s0} +\tilde{k}_s)^2}\bigg)
             \bigg(1 +\frac{1}{2}\frac{|\textbf{q}_i|^2}{(k_{i0} +\tilde{k}_i)^2}\bigg)\bigg(1 +\frac{1}{2}\frac{|\textbf{q}_s+\textbf{q}_i|^2}{(k_{p0}+\tfrac{v_{gs}}{v_{gp}} \tilde k_s +\tfrac{v_{gi}}{v_{gp}} \tilde k_i)^2}\bigg)\\ 
             &\quad\times \text{sinc}\biggl(\frac{L_z}{2}\bigg(\tilde k_s+\tilde k_i-\bigg(\frac{v_{gs}}{v_{gp}} \tilde k_s +\frac{v_{gi}}{v_{gp}} \tilde k_i\bigg)-\frac{1}{2}\biggl[\frac{|\textbf{q}_s|^2}{  k_{s0} +\tilde{k}_s}
             +\frac{|\textbf{q}_i|^2}{  k_{i0} +\tilde{k}_i} 
             - \frac{|\textbf{q}_s+\textbf{q}_i|^2}{k_{p0}+\tfrac{v_{gs}}{v_{gp}} \tilde k_s +\tfrac{v_{gi}}{v_{gp}} \tilde k_i}\biggr] \biggr) \biggr) \\
             &\quad \times \alpha_p(\textbf{q}_s+\textbf{q}_i) \tilde \alpha_p ( \tfrac{v_{gs}}{v_{gp}} \tilde k_s +\tfrac{v_{gi}}{v_{gp}} \tilde k_i) \,  \hat{a}_s^{\dagger}(\textbf{k}_s) \hat{a}_i^{\dagger}(\textbf{k}_i) + h.c.\label{eq:spdc}
    \end{split}
    \end{equation}
Even for relatively broad-band fields we have $\tilde k \ll k_0$ for pump, signal and idler, so we can approximate $\tilde k + k_0\approx k_0$ in the Sinc quotients, and for close to co-linear generation $\frac{|\textbf{q}_i|^2}{(k_{i0} +\tilde{k}_i)^2}\ll1$, so we can neglect polynomial terms in front of the Sinc of the order $\mathcal{O}(|\textbf{q}|^2/k_0^2)$ and above, to obtain,
\begin{equation}
    \begin{split}
             \int H^{NL}(t) dt &\propto \int d^2\textbf{q}_i d \tilde k_i\int d^2\textbf{q}_s d \tilde k_s \textrm{ } \text{sinc}\biggl(\frac{L_z}{2}\bigg(\tilde k_s+\tilde k_i-\bigg(\frac{v_{gs}}{v_{gp}} \tilde k_s +\frac{v_{gi}}{v_{gp}} \tilde k_i\bigg)-\frac{1}{2}\biggl[\frac{|\textbf{q}_s|^2}{  k_{s0}}
             +\frac{|\textbf{q}_i|^2}{  k_{i0}} 
             - \frac{|\textbf{q}_s+\textbf{q}_i|^2}{k_{p0}}\biggr] \biggr) \biggr) \\
             &\quad \times \alpha_p(\textbf{q}_s+\textbf{q}_i)\tilde \alpha_p(\tfrac{v_{gs}}{v_{gp}} \tilde k_s +\tfrac{v_{gi}}{v_{gp}} \tilde k_i) \,  \hat{a}_s^{\dagger}(\textbf{k}_s) \hat{a}_i^{\dagger}(\textbf{k}_i) + h.c.\label{eq:spdc}
    \end{split}
    \end{equation}
    
At degeneracy, where $\tilde k_s=\tilde k_i=0$, we have,
\begin{equation}
    \begin{split}
             \int H^{NL}(t) dt &\propto \int d^2\textbf{q}_i d \tilde k_i\int d^2\textbf{q}_s d \tilde k_s \textrm{ } \text{sinc}\biggl(\frac{L_z}{4}\biggl[\frac{|\textbf{q}_s+\textbf{q}_i|^2}{k_{p0}}-\frac{|\textbf{q}_s|^2}{  k_{s0}}
             -\frac{|\textbf{q}_i|^2}{  k_{i0} } 
             \biggr] \biggr) \times \alpha_p(\textbf{q}_s+\textbf{q}_i) \,  \hat{a}_s^{\dagger}(\textbf{k}_s) \hat{a}_i^{\dagger}(\textbf{k}_i) + h.c.\label{eq:spdcapprox}
    \end{split}
    \end{equation}
    To simplify further, we define a scaled transverse momenta,
\begin{equation}
\begin{split}
     \tilde{\textbf{q}}_s &:= \left( \frac{k_{p0}}{k_{s0}}-1 \right)^{\tfrac{1}{2}}\textbf{q}_s :=c_s \textbf{q}_s \,
\end{split}
\end{equation}  
and analogously for the idler. We write $\varepsilon:=(1-\frac{1}{c_s c_i})$ and express the product of the Sinc function and pump profile $\alpha_p$ as:
\begin{equation}
     F(\textbf{q}_s,\textbf{q}_i)\propto\text{sinc}\biggl( \frac{L_z}{ 4k_{p0}} \biggl[(2+\varepsilon)|\tilde{\textbf{q}}_-|^2+\varepsilon |\tilde{\textbf{q}}_+|^2\biggr] \biggr) \times \alpha_p({\textbf{q}}_s+{\textbf{q}}_i),
\end{equation}
where $|\tilde{\textbf{q}}_{\pm}|:=\tfrac{\left|c_s \textbf{q}_s\pm c_i \textbf{q}_i\right|}{\sqrt{2}}$. We define the above expression as the \textit{joint-transverse-momentum-amplitude} (JTMA). 

For $\lambda_s=1550$nm, we have $c_s=0.9677$, $c_i=1.0139$ and for $\lambda_s=810$nm, $c_s=0.9979$, $c_i=1.047$. We approximate $\varepsilon\approx 0$ because $\varepsilon=-0.0193$ at 1550nm and $\varepsilon=0.043$ at 810nm. Under this approximation, the JTMA can be simplified to:
\begin{equation}
\begin{split}
     F(\textbf{q}_s,\textbf{q}_i)\propto \text{sinc}\biggl(\frac{L_z}{2 k_{p0}} |\tilde{\textbf{q}}_-|^2 \biggr)\times \alpha_p({\textbf{q}}_s+{\textbf{q}}_i).
\end{split}
\end{equation}
We see that colinear Type-II phase-matching with periodic poling deviates somewhat from the idealised JTMA in the main text (Eq.~\ref{eq:JTMAformula}) owing to the non-unit $c_s$, $c_i$, and the non-vanishing $\varepsilon$. However, this discrepancy becomes increasingly benign for the collection-limited systems in consideration in which the Sinc contribution becomes dominated instead by the collection optics.

Rewriting the normalised JTMA for $c_s = c_i \approx 1$ (and remaining at degeneracy $\lambda_s = \lambda_i$), we impose a Gaussian transverse pump profile across the crystal, to obtain
\begin{equation}
    F(\textbf{q}_s,\textbf{q}_i) = \mathcal{N}_1\textrm{sinc}\biggl(\frac{|\textbf{q}_s - \textbf{q}_i|^2}{\sigma_S^2} \biggr) \times \textrm{exp}\biggl(-\frac{|\textbf{q}_s + \textbf{q}_i|^2}{2\sigma_P^2} \biggr),
\end{equation}
where $\sigma_S = \sqrt{4 k_{p0}/L_z}$ and $\mathcal{N}_1$ is a normalisation constant. Here the transverse pump momentum profile $\alpha_P(\textbf{q}_p)$ is a Gaussian, where $w_P$ is the pump radius in position space, so that $\sigma_P = \frac{\sqrt{2}}{w_P}$. 

Transforming to the Fourier plane with 2$f$-lens configuration, we have ($\textbf{q}\rightarrow \frac{2 \pi\textbf{x}}{f \lambda}$),
\begin{equation}
\begin{split}
F\left(\textbf{x}_s,\textbf{x}_i\right)\propto \text{sinc}\biggl(\frac{1}{\sigma_S^2} \left|\frac{2 \pi}{f \lambda_s}{\textbf{x}}_s- \frac{2 \pi}{f \lambda_i}{\textbf{x}}_i\right|^2\biggr) \alpha_p\biggl(-\frac{1}{2 \sigma_P^2}\left|\frac{2 \pi}{f \lambda_s}{\textbf{x}}_s + \frac{2 \pi}{f \lambda_i}{\textbf{x}}_i\right|^2\biggr).
\end{split}
\end{equation}

\section{Collected JTMA to Collection Limited JTMA}\label{sec:Collected_to_CL}
\begin{figure*}[ht!]
\centering\includegraphics[width=0.5\textwidth]{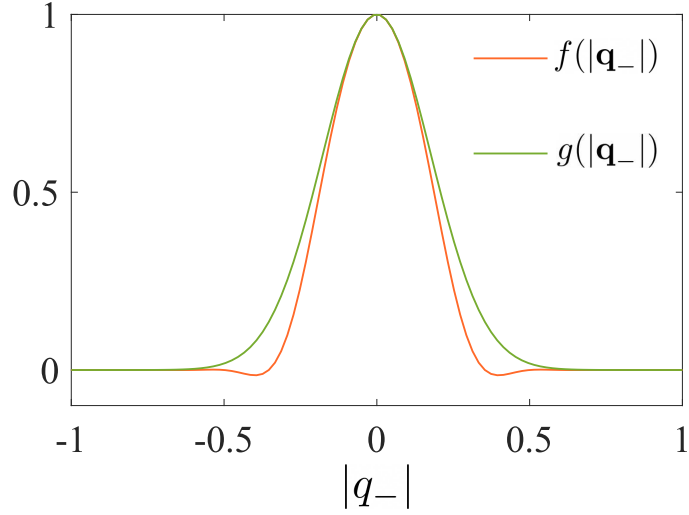} 
\caption{\textbf{Comparison between product of Sinc and Gaussian term ($f(|\textbf{q}_-|)$) and the collection Gaussian envelope ($g(|\textbf{q}_-|)$)}: For $\sigma_S \gtrapprox\sqrt{2}\sigma_C$, the product of the Sinc term and Gaussian envelope can be approximated by a Gaussian with width $\sigma_C$}
\label{fig:CollectedCL}
\end{figure*}
We introduce the effect of the collection mode in $F(\textbf{q}_s,\textbf{q}_i)$ by considering collected JTMA $G(\textbf{q}_+,\textbf{q}_-)$ given in Eq.~\eqref{eq:collectedJTMASumDiff}. To approximate the collected JTMA to collection limited JTMA, we investigate the approximation of the product of Sinc and the collection Gaussian envelope:
\begin{align}
\begin{split}
    f(|\textbf{q}_-|) =  \mathcal{N}\exp\biggl\{\frac{-|\textbf{q}_-|^2}{\sigma_C^2}\biggr\}\textrm{sinc}\biggl(\frac{2 |\textbf{q}_-|^2}{\sigma_S^2}\biggr)\,,
\end{split}
\end{align}
with $\mathcal{N}$ a normalisation factor, by the Gaussian function,
\begin{align}
\begin{split}
     g(|\textbf{q}_-|) = \mathcal{N}'\exp\biggl\{\frac{-|\textbf{q}_-|^2}{\sigma_C^2}\biggr\}\,.
\end{split}
\end{align}
These terms have inner product, 
\begin{align}
\begin{split}
     \int d^2 \textbf q_- \mathcal{N}\exp\biggl\{\frac{-|\textbf{q}_-|^2}{\sigma_C^2}\biggr\}\textrm{sinc}\biggl(\frac{2|\textbf{q}_-|^2}{\sigma_S^2}\biggr) \mathcal{N}'\exp\biggl\{\frac{-|\textbf{q}_-|^2}{\sigma_C^2}\biggr\}\geq 0.99
\end{split}
\end{align}
for $\sigma_S > 1.4161\sigma_C\gtrapprox \sqrt{2}\sigma_C$. Slightly looser approximation sees an inner product of 0.95 achieved at $\sigma_S >\tfrac{3}{2\sqrt{2}} \sigma_C$.

Furthermore, in the case of calculating the singles count rates in Eq.~\eqref{eq:singlePhotonStatisticsMain}, in which only the herald photon imparts a collection envelope we have a mildly stronger condition for the approximation to hold, 
\begin{align}
\begin{split}
     \mathcal{N}\exp\biggl\{\frac{-|\textbf{q}_-|^2}{2\sigma_C^2}\biggr\}\textrm{sinc}\biggl(\frac{2 |\textbf{q}_-|^2}{\sigma_S^2}\biggr)\approx  \mathcal{N}'\exp\biggl\{\frac{-|\textbf{q}_-|^2}{2\sigma_C^2}\biggr\}\,,
\end{split}\label{eq:sincprodgauss}
\end{align}
leading to an additional factor of $\sqrt{2}$ arising in the equivalent conditions above, ie. $\sigma_S \gtrapprox 2\sigma_C$ implies greater than 0.99 inner product. 

\section{Calculations of $\text{Pr}(a,-a)$ and $\text{Pr}(a,a)$}\label{sec:theorycal}
\begin{figure*}[ht!]
\centering\includegraphics[width=1\textwidth]{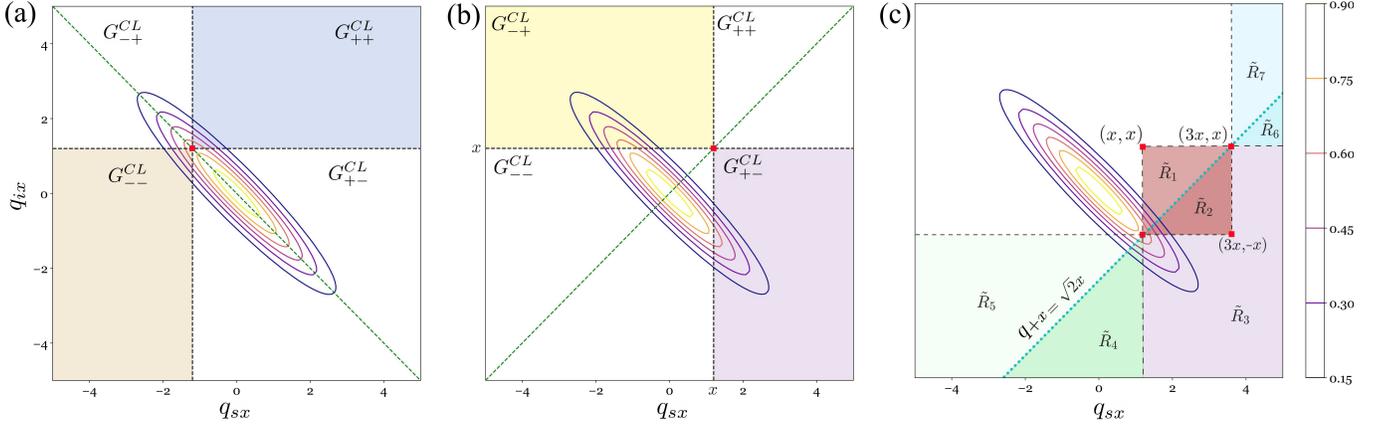} 
\caption{\textbf{Illustration of regions in collected JTMA to approximate $\text{Pr}(a,-a)$ and $\text{Pr}(a,a)$}: (a) For $\text{Pr}(a,-a)$, the shaded regions $G^{CL}_{++}$ and $G^{CL}_{--}$ are equal due to the symmetry about $q_{+x}=0$  (green, dashed). (b) Similarly, for $\text{Pr}(a,a)$, the regions $G^{CL}_{-+}$ and $G^{CL}_{+-}$ are equal due to symmetry across $q_{-x} = 0$. In (c) $G^{CL}_{+-} = \tilde{R}_1+\tilde{R}_2 +\tilde{R}_3$, which is calculated from various integrals (See Eq.~\eqref{eq:C4} and \eqref{eq:C5})}
\label{fig:regionplots}
\end{figure*}
For convenience, the expression of two photon coincidence probability in Eq.~\eqref{eq:twophotonprob} can be written in the reduced form
\begin{align}\label{eq:2dpiregion}
   \text{Pr}(a_s, a_i) = \bigl|G_{++}+ G_{--} - G_{-+} - G_{+-}\bigr|^2.
\end{align}{}
For the signal photon, the subscript $+$ represents the integration performed in the interval $[a_s,\infty)$ and $-$ corresponds to the integration in the interval $(-\infty,a_s]$ (same for idler). 

To calculate $\text{Pr}(a,-a)$, we expand the integral for $\text{Pr}(a,-a)$ in terms of integral by regions $G^{CL}_{++}$, $G^{CL}_{--}$, $G^{CL}_{-+}$, and $G^{CL}_{+-}$ (see Fig.~\ref{fig:regionplots}a). From symmetry, we have $G^{CL}_{--}$ = $G^{CL}_{++}$, and with Eq.~\eqref{eq:2dpiregion} we write $\text{Pr}(a,-a)$ as,
\be
\text{Pr}(a,-a) = |N - 4G^{CL}_{--}|^2,
\label{eq:praastep1}
\ee
where $N = G^{CL}_{--} + G^{CL}_{++} +G^{CL}_{-+} +G^{CL}_{+-}=\pi\sqrt{2\pi}\mathcal{N}_1\frac{\sigma_C^3\sigma_P^2}{\sigma_C^2 + \sigma_P^2}$. Assuming $\sigma_P\ll\sigma_C$ in the bow-tie region formed by $G^{CL}_{--}$ and $G^{CL}_{++}$, the factor $\exp\bigl\{-\frac{(q_{sx}-q_{ix})^2}{2\sigma_C^2}\bigr\}$ is approximated by $\exp\big(-\frac{2a^2}{\sigma_C^2}\big)$, which reduces the expression in Eq.~\eqref{eq:praastep1} to
\begin{align}\label{eq:pxmx1}
\text{Pr}(a,-a) &\approx \bigg|N - 4 \int_{-\infty}^{a}\int_{-\infty}^{-a}\mathcal{G}^{DG}(q_{sx},q_{ix};a) \text{d} q_{sx}\text{d} q_{ix}\bigg|^2
\end{align}
where 
\be
\mathcal{G}^{DG}(q_{sx},q_{ix};a)  \propto  \exp \bigg(-\frac{(q_{sx}+q_{ix})^2}{2\tilde{\sigma}_P^2}\bigg)\exp\bigg(-\frac{2a^2}{\sigma_C^2}\bigg),
\ee
for which there exist a simple analytic solution,
\be
\begin{split}
\Pr(a,-a)&\approx  \left|N - N'\exp\bigg(-\frac{2a^2}{\sigma_C^2}\bigg)\right|^2 \\
N'&:=\sqrt{2\pi}\mathcal{N}_1\frac{ \pi  \sigma _C^3 \sigma _P^3}{\left(\sigma _C^2+\sigma _P^2\right)^{3/2}}
\end{split}
\ee
given in Eq.~\eqref{eq:pxmxclosed}. One can write the expression for the visibility $V$ in terms of $N$ and $N'$ as
\be\label{eq:visbilityinpraa}
    V = \frac{|N|^2 - |N-N'|^2}{|N|^2 + |N-N'|^2} \approx \frac{\sigma_C^2 - |\sigma_C - \sigma_P|^2}{\sigma_C^2 + |\sigma_C - \sigma_P|^2}\quad\textrm{ for } \sigma_P\ll \sigma_C.
\ee
For a fixed value of $\sigma_C$, obtained from the fit of $\Pr(a,-a)$, one can also get the pump width parameter $\sigma_P$ from the Eq.~\eqref{eq:visbilityinpraa}, given the visibility of the experimental data. 

Therefore, for $\text{Pr}(a,a)$, we follow the same steps as $\text{Pr}(a,-a)$ of writing the complete integral as the sum of integrals over several regions (see Fig.~\ref{fig:regionplots}b). Here, the integrals $G^{CL}_{-+}$ and $G^{CL}_{+-}$ are equal ($G^{CL}_{+-} = G^{CL}_{-+}\equiv \tilde{R}$). We aim to partition the space into areas which have closed-form integral solutions and a bow-tie region which can be well approximated under the previous reasoning. We expand $\tilde{R} = \tilde{E} + F - (\tilde{F}_1 + \tilde{F}_2)$, where the contributions are given as,
\be
\begin{split}\label{eq:C4}
    \tilde{E}  &= \iint_{\mathcal{L}} G^{CL}(q_{+x},q_{-x}) \text{d} q_{+x}\text{d} q_{-x}, \qquad F = \frac{1}{2}\int_{a}^{3a} \int_{-a}^{a}  \mathcal{G}^{DG}(q_{sx},q_{ix};a)   \text{d} q_{sx}\text{d} q_{ix},  \\
    \tilde{F}_1 &= \frac{1}{2}\int^{a}_{-\infty} \int^{-a}_{-\infty} \mathcal{G}^{DG}(q_{sx},q_{ix};a) \text{d} q_{sx}\text{d} q_{ix}, \qquad \tilde{F}_2 = \frac{1}{2}\int^{\infty}_{3a} \int^{\infty}_{a} \mathcal{G}^{DG}(q_{sx},q_{ix};a)\text{d} q_{sx}\text{d} q_{ix},
\end{split}
\ee
where $G^{CL}(q_{+x},q_{-x})$ is integrated over the region $\mathcal{L}\in(q_{-x}\leq\sqrt{2}x)$.
The rationale behind this expansion can be seen when we further express each terms in the sum of $\tilde{R}$ into the regions shown in Fig.~\ref{fig:regionplots}c,
\be
\begin{split}\label{eq:C5}
    \tilde{E}  &= \tilde{R}_2 + \tilde{R}_3 + \tilde{R}_4 +\tilde{R}_6, \qquad F = \frac{1}{2}(\tilde{R}_1 +\tilde{R}_2)= \tilde{R}_1, \\
    \tilde{F}_1 &= \frac{1}{2}(\tilde{R}_5  + \tilde{R}_4) = \tilde{R}_4, \qquad \tilde{F}_2 = \frac{1}{2}(\tilde{R}_7  +\tilde{R}_6) = \tilde{R}_6, \\
    \tilde{R} &= \tilde{E} + F - (\tilde{F}_1 + \tilde{F}_2) = \tilde{R}_1+\tilde{R}_2 +\tilde{R}_3.
\end{split}
\ee
We solve for $N - 4\tilde{R}$ with the help of integral tables given for error functions and the fact that $\text{Pr}(a,a)$ is an even function, and we get,
\begin{equation}
    N - 4\tilde{R} =  \mathcal{A}\biggl[ \sqrt{2} \sigma_P\text{ } e^{-2a^2\big(\frac{1}{\tilde{\sigma}_P^2} + \frac{1}{\sigma_C^2}\big)} - 2\sqrt{\pi}\text{ }|a|\text{ } e^{\frac{-2a^2}{\sigma_C^2}}  \times\text{ erfc}\bigg(\frac{\sqrt{2}|a|}{\tilde{\sigma}_P}\bigg) -\frac{\pi\sigma_C}{2\sqrt{2}}\bigg\{1-2\text{erf}\bigg(\frac{\sqrt{2}|a|}{\sigma_C}\bigg)\bigg\} \biggr].\label{eq:pxx12}
\end{equation}

To fit $\text{Pr}(a,-a)$ and $\text{Pr}(a,a)$, one requires the estimated location of origin from the experiment data, which we can obtain from 
\be
\textrm{Pr}(a,-\infty) \propto \abs{\textrm{erf}\bigg(\frac{\sqrt{2}(a-a_s)}{\sigma_C}\bigg)}^2, \textrm{Pr}(-\infty,a) \propto \abs{\textrm{erf}\bigg(\frac{\sqrt{2}(a-a_i)}{\sigma_C}\bigg)}^2,
\ee
where $\text{Pr}(a,-\infty)$ is minimum for $a = a_s$ ($\text{Pr}(-\infty,a)$ is minimum for $a = a_i$). Hence, the location of origin coincides with the location of the respective minimas ($q_{sx} = a_s, q_{ix} = a_i$). 

\section{Mode Basis Design and Heralding Efficiency Details}\label{sec:DesginApp}
\begin{figure}[ht!]
\centering\includegraphics[width=0.5\textwidth]{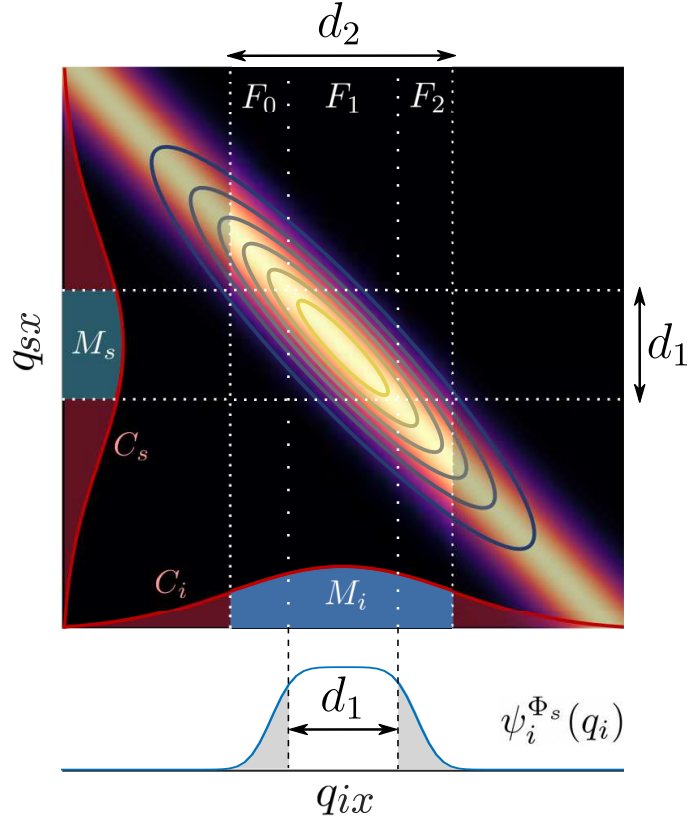}
\caption{Heralding of the idler photon state by projection onto a signal photon. The signal photon is projected onto the measurement mode, $M_s(\textbf q_s) = \Phi_s(\textbf q_s)\mathcal{C}(\textbf q_s|\sigma_C)$ (turquoise, left axis) determined by the collection mode, $\mathcal{C}_s$ (red, left axis) and a pixel hologram $\Phi_s$, of diameter $d_1$. The resultant heralded idler photon state, $\psi^{\Phi_s}_i(\textbf q_i)$, (blue, bottom), is determined by the projection of $M_s(\textbf q_s)$ onto the JTMA (contours), see Eq.~\eqref{eq:singlePhotonStatistics}, which owing to the relative size of $\sigma_C$, can be approximated by the correlation term (density plot) (see Appendix~\ref{sec:Collected_to_CL}). The finite width of the JTMA leads to contributions arising from regions $F_0$ and $F_2$ that result in a heralded photon with a width larger than $d_1$. The idler measurement mode, $M_i(\textbf q_i) = \Phi_i(\textbf q_i)\mathcal{C}(\textbf q_i|\sigma_C)$ (blue region, lower plot axes), is choosen to have an increased pixel diameter $d_2>d_1$, leading to an increased coincidence probability given by the inner product, $\langle M_i^{\Phi_s},\psi^{\Phi_s}_i \rangle$ (Eq.~\eqref{eq:generalVectorStatisticsOnHerald}), whilst the (inclusive) singles probability of the signal detection is $\langle \psi^{\Phi_s}_i,\psi^{\Phi_s}_i \rangle$ (Eq.~\eqref{eq:singlePhotonStatistics}) independent of $d_2$. Hence, an increased $d_2$ leads to an increase in the one-sided heralding efficiency (Eq.~\eqref{eq:heraldingEffi}).} 
\label{fig:heraldingeff}
\end{figure}

To produce a postselected state experimentally, we project holograms $\{\Phi_s(\textbf{q}_s)\}_a$ and $\{\Phi_i(\textbf{q}_i)\}_b$  on the SLMs with an additional constraint $|\Phi_s(\textbf{q}_s)|\leq1$. This results in a state characterised by the complex elements, $T_{ab}$, given by, 
\begin{equation}
    T_{ab} =\int d^2\textbf{q}_{s}\int d^2\textbf{q}_{i} \Phi_{s}^a(\textbf{q}_{s}) \Phi_{i}^b(\textbf{q}_{i}) G(\textbf{q}_{s},\textbf{q}_{i}).
\end{equation}
describing the subnormalised postselected state, $\ket{\psi^{(PS)}}$, which is defined on the corresponding normalised discrete mode basis $\{\ket{\hat{e}^a_s}\}_a$,
\begin{equation}
\begin{split}
\label{eq:generalModeBasisState}
\ket{\psi^{(PS)}}&:=  \sum_{ab} T_{ab}  \ket{ \hat e_{s}^a} \ket{ \hat e_{i}^b},
\end{split}
\end{equation}
where, to ensure $\{\ket{\hat{e}^a_s}\}_a$ are normalised, we write
\begin{equation}
    |\braket{\hat e_s^a}{ \hat e_s^{a}}|^2=1 =N^a_s\bigg|\int d^2\textbf{q}_{s} \Phi_{s}^a(\textbf{q}_{s})\mathcal{C}(\textbf q_s|\sigma_C)\bigg|^2 =N^a_s\bigg|\int d^2\textbf{q}_{s} M^a_s(\textbf{q}_{s})\bigg|^2
\end{equation}
with $N_s^a$ as normalisation factors (similar for idler). We construct the holograms so that $\ket{\hat e_{n}^a}$ ($n=s,i$) are orthogonal and form the discrete mode basis, which can easily be achieved by for instance making $\{\Phi_s(\textbf{q}_s)\}_a$ disjoint (pixels). The resultant state can be understood as the full biphoton state filtered through the collection mode apertures defined by $\hat L_s$ and $\hat L_i$,
\begin{align}
\begin{split}
\label{eq:CollectionFiltering}
\hat L_s &:= \sum_a \frac{1}{\sqrt{N^a_s}} \ketbra{\hat e^a_s}{\hat e^a_s}\\
\ket{\psi^{(PS)}}&:=\hat L_s \hat L_i \ket {\psi^{(bi)}}.\\
\end{split}
\end{align}
When measuring arbitrary vectors in this subspace, for instance $\ket{\vec v^s} = \sum_a v^s_a \ket{\hat e^a}$, with $|\langle \vec v^s | \vec v^s\rangle | =1$, one constructs the holograms as
\begin{equation}
\begin{split}
\label{eq:generalVectorState}
\Phi^{\vec v}_s(\textbf q_s) = A^{\vec v} \sum_a v_a \Phi^a_s(\textbf q_s)
\end{split}
\end{equation}
with $A^{\vec v}$ chosen so that $\max_{\textbf q} |\Phi^{\vec v}_s(\textbf q)| \leq 1$. These holograms result in the measurement statistics,

\begin{equation}
\begin{split}
\label{eq:generalVectorStatistics}
\Pr(\vec{v}^s,\vec{v}^i)&= \left(A^{\vec{v}^s}A^{\vec{v}^i}\right)^2\left| \bra{\vec v^s}\bra{\vec v^i}\ket{\psi^{(PS)}} \right|^2\\
\end{split}
\end{equation}
The choice of $A^{\vec v}$ results in an effective change in the postselection probability. 
Similarly, probability of obtaining a (inclusive) single signal photon when measuring these states is 
\begin{equation}
\begin{split}
\label{eq:singlePhotonStatistics}
 \psi^{\Phi_s}_i (\textbf q_i) &:= \int d^2 \textbf{q}_s\Phi^{\vec v^s}_s(\textbf q_s)\mathcal{C}(\textbf q_s|\sigma_C) F(\textbf q_s,\textbf q_i) \\
\Pr(\vec{v}^s)&= \int d^2 \textbf{q}_i\left| \psi^{\Phi_s}_i (\textbf q_i) \right|^2\\
&=\left(A^{\vec{v}^s}\right)^2 \int d^2 \textbf{q}_i\left| \bra{\vec v^s}\bra{\textbf{q}_i} \hat L_s\ket{\psi^{bi}} \right|^2\\
\end{split}
\end{equation}
with $\psi^{\Phi_s}_i (\textbf q_i)$ describing the pure heralded idler photon state after heralding with $\Phi^{\vec v^s}_s$ on the signal. The coincidence probability can be written similarly in terms of this heralded idler photon state,
\begin{equation}
\begin{split}
\label{eq:generalVectorStatisticsOnHerald}
\Pr(\vec{v}^s,\vec{v}^i)&= \left| \int d^2 \textbf{q}_i M^{\Phi_s}_i(\textbf q_i) \psi^{\Phi_s}_i (\textbf q_i) \right|^2\\
\end{split}
\end{equation}

The one-sided heralding efficiency is given by 
\begin{equation}
\begin{split}
\eta^{s\rightarrow i}=\frac{\Pr(\vec{v}^s,\vec{v}^i)}{\Pr(\vec{v}^s)}
\end{split}
\end{equation}
which can be optimised by ensuring the idler mode basis has large overlap with the heralded photons from the signal, as well as choosing measurements for which $A^{\vec v_i}$ may be large. In Fig.~\ref{fig:heraldingeff} we depict the increased heralding efficiency associated to increasing the size of one party's pixel relative to the other. 

\end{document}